\documentclass[aps,pra,reprint,superscriptaddress]{revtex4-1}
\usepackage{amsmath}

\usepackage{resizegather}

\usepackage{amssymb}
\usepackage{physics}
\usepackage{braket}
\usepackage{dsfont}
\usepackage{mathtools}
\usepackage{diagbox}
\usepackage{varwidth}
\usepackage{framed}
\usepackage{natbib}
\usepackage{soul,xcolor} 

\usepackage{epstopdf}
\usepackage{float}

\usepackage{tikz} 
\usetikzlibrary{decorations.pathmorphing,patterns}
\usetikzlibrary{backgrounds}
\usetikzlibrary{decorations.markings}
\tikzset{
    set arrow inside/.code={\pgfqkeys{/tikz/arrow inside}{#1}},
    set arrow inside={end/.initial=>, opt/.initial=},
    /pgf/decoration/Mark/.style={
        mark/.expanded=at position #1 with
        {
            \noexpand\arrow[\pgfkeysvalueof{/tikz/arrow inside/opt}]{\pgfkeysvalueof{/tikz/arrow inside/end}}
        }
    },
    arrow inside/.style 2 args={
        set arrow inside={#1},
        postaction={
            decorate,decoration={
                markings,Mark/.list={#2}
            }
        }
    },
}
\usetikzlibrary{shapes}
\definecolor{myred}{HTML}{FF0000} 
\definecolor{QRc}{HTML}{E64F40} 
\definecolor{nonQRc}{HTML}{3C49E3} 
\definecolor{eprsource}{rgb}{0.34, 0.81, 0.54}

    \usepackage{xcolor}
\usepackage{mdframed}
\usepackage{multienum}

\makeatletter
\newcommand*{\centerfloat}{%
  \parindent \z@
  \leftskip \z@ \@plus 1fil \@minus \textwidth
  \rightskip\leftskip
  \parfillskip \z@skip}
\makeatother

\newcommand{\san}[1]{\mathsf{#1}}

\usepackage{todonotes}

\usepackage{hyperref}

\newcommand{\nstar}{n^{\star}}
\newcommand{\proj}[1]{|#1\rangle\langle#1|}

\usetikzlibrary{shapes.geometric}
\usetikzlibrary{patterns}
\usetikzlibrary{positioning}
\usepackage[T1]{fontenc}
\usepackage{lmodern}
\usepackage{notoccite}
\usepackage{amssymb}

\tikzset{
    buffer/.style={
        draw,
        shape border rotate=180,
        regular polygon,
        regular polygon sides=3,
        fill=gray, fill opacity = 0.2,
        node distance=2cm,
        minimum height=4em
    }
}

\usepackage[noend]{algpseudocode}
\usepackage{algorithm}
\algnewcommand{\Initialize}[1]{%
  \State \textbf{Initialize:}
  \Statex \hspace*{\algorithmicindent}\parbox[t]{.8\linewidth}{\raggedright #1}
}

\makeatletter
\def\BState{\State\hskip-\ALG@thistlm}
\makeatother

\setstcolor{red} 

\begin{document}
\title{Parameter regimes for a single sequential quantum repeater \vspace{-2mm}}
\author{F. Rozp\k{e}dek}
\thanks{These authors contributed equally}
\email{f.d.rozpedek@tudelft.nl}
\affiliation{QuTech, Delft University of Technology, Lorentzweg 1, 2628 CJ Delft, The Netherlands}
\author{K. Goodenough}
\thanks{These authors contributed equally}
\email{f.d.rozpedek@tudelft.nl}
\affiliation{QuTech, Delft University of Technology, Lorentzweg 1, 2628 CJ Delft, The Netherlands}
\author{J. Ribeiro}
\affiliation{QuTech, Delft University of Technology, Lorentzweg 1, 2628 CJ Delft, The Netherlands}
\author{N. Kalb}
\affiliation{QuTech, Delft University of Technology, Lorentzweg 1, 2628 CJ Delft, The Netherlands}
\affiliation{Kavli Institute of Nanoscience, Delft University of Technology, Lorentzweg 1, 2628 CJ Delft, The Netherlands}
\author{V. Caprara Vivoli}
\affiliation{QuTech, Delft University of Technology, Lorentzweg 1, 2628 CJ Delft, The Netherlands}
\author{A. Reiserer}
\affiliation{QuTech, Delft University of Technology, Lorentzweg 1, 2628 CJ Delft, The Netherlands}
\affiliation{Kavli Institute of Nanoscience, Delft University of Technology, Lorentzweg 1, 2628 CJ Delft, The Netherlands}
\affiliation{Quantum Networks Group, Max-Planck-Institute of Quantum Optics, Hans-Kopfermann-Str.~1, 85748 Garching, Germany}
\author{R. Hanson}
\affiliation{QuTech, Delft University of Technology, Lorentzweg 1, 2628 CJ Delft, The Netherlands}
\affiliation{Kavli Institute of Nanoscience, Delft University of Technology, Lorentzweg 1, 2628 CJ Delft, The Netherlands}
\author{S. Wehner}
\affiliation{QuTech, Delft University of Technology, Lorentzweg 1, 2628 CJ Delft, The Netherlands}
\author{D. Elkouss}
\affiliation{QuTech, Delft University of Technology, Lorentzweg 1, 2628 CJ Delft, The Netherlands}
\begin{abstract}
Quantum key distribution allows for the generation of a secret key between distant parties connected by a quantum channel such as optical fibre or free space. 
Unfortunately, the rate of generation of a secret key by direct transmission is fundamentally limited by the distance. This limit can be overcome by the implementation of so-called quantum repeaters. 
Here, we assess the performance of a specific but very natural setup called a single sequential repeater for quantum key distribution. We offer a fine-grained assessment of the repeater by introducing a series of \emph{benchmarks}. 
The benchmarks, which should be surpassed to claim a working repeater, are based on finite-energy considerations, thermal noise and the losses in the setup. 
In order to boost the performance of the studied repeaters we introduce two methods. The first one corresponds to the concept of a \emph{cut-off}, which reduces the effect of decoherence during storage of a quantum state by introducing a maximum storage time. Secondly, we supplement the standard classical post-processing with an \emph{advantage distillation} procedure.
Using these methods, we find realistic parameters for which it is possible to achieve rates greater than each of the benchmarks, guiding the way towards implementing quantum repeaters.
\end{abstract}    
    \maketitle
\onecolumngrid
\section{Introduction}

Quantum communication enables the implementation of tasks with qualitative advantages with respect to classical communication, including secret key distribution~\cite{Bennett_84,Ekert_91}, clock synchronization~\cite{Giovanetti_01} and anonymous state transfer~\cite{christandl2005quantum}. 
Unfortunately, the transmission of both classical and quantum information over optical fibres decreases exponentially with the distance. 
While the problem of losses applies both to classical and quantum communication, classical information can be amplified at intermediate nodes, preventing the signal from dying out and thus increasing the rate of transmitted information. 
At the same time, the existence of a quantum analogue of a classical amplifier is prohibited by the no-cloning theorem~\cite{wootters1982single}. Fortunately, in principle it is possible to construct a \emph{quantum repeater} to increase the rate of transmission without having to amplify the signal~\cite{briegel1998quantum, Munro_15}. Hence, the construction of a quantum repeater would represent a fundamental milestone towards long-distance quantum communications.

The basic idea of a quantum repeater protocol has undergone many changes since its original proposal~\cite{briegel1998quantum}. The authors of this scheme showed that by dividing the entire communication distance into smaller segments, generating entanglement over those short links and performing entanglement swapping operation at each of the intermediate nodes in a nested way, one can establish long-distance entanglement. It was also shown that by including the procedure of entanglement distillation, one can furthermore overcome the problem of noise. Effectively, the authors proposed a scheme that enables to generate a high-fidelity long-distance entangled link with an overhead in resources that scales polynomially with distance. Unfortunately, this model does not go into detail of how the physical imperfections of realistic devices, such as decoherence of the quantum memories with time or possibly the probabilistic nature of entanglement swapping, affect the performance. These observations have led to the development of significantly more detailed and accurate, but at the same time significantly more complex, repeater schemes~\cite{duan2001long, jiang2009quantum, munro2010quantum, munro2012quantum, Azuma_15}. Many quantum repeater proposals require significant resources and are thus not within experimental reach. However, the recent experimental progress in the development of quantum memories~\cite{reiserer2016robust, lvovsky2009optical, specht2011single} has brought the realisation of a quantum repeater closer than ever.

\begin{center}
\begin{figure}
\includegraphics[scale=1.05,clip,trim = 60mm 209mm 60mm 19mm]{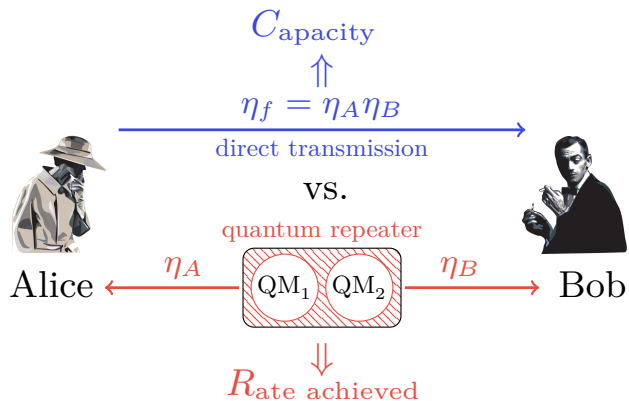}
\caption{The quantum repeater will send photons entangled with the $\textrm{QM}_1$ to Alice through the optical fibre of transmissivity $\eta_A$. After receiving one photon she will perform a BB84 or six-state measurement. After Alice has measured a photon and communicated her success to the quantum repeater, the quantum repeater tries to send a photon entangled with the $\textrm{QM}_2$ to Bob through the optical fibre of transmissivity $\eta_B$. If Bob does not receive a photon within some pre-defined amount of trials (i.e.~the cut-off), Alice and Bob will abort the round.
This is done to prevent the state in the $\textrm{QM}_1$ from decohering excessively. If Bob does succeed, the quantum repeater performs a Bell state measurement on the two quantum memories.}
\label{fig:modelofsetup}
\end{figure}
\end{center}

In this paper, we evaluate a realistic setup of a so-called single sequential quantum repeater on how it performs for the specific task of quantum key distribution. Our interest in assessing the repeater with respect to this task is motivated by the fact that quantum key distribution is, at the moment, the most mature quantum technology~\cite{scarani2009security}. 
The setup considers two parties which we call Alice and Bob who are spatially separated, and want to generate a shared secret key. The setup that we will investigate here was originally proposed in~\cite{luong2015overcoming}, where the authors were inspired by the memory-assisted measurement-device-independent QKD setup (MA-MDI QKD)~\cite{panayi2014memory}.
Alice and Bob use a single sequential quantum repeater located between them, where both of them are connected to the quantum repeater by optical fibre. The repeater is composed of two quantum memories, both of which have the ability to become entangled with a photon, see FIG.~\ref{fig:modelofsetup}. However, the repeater has a single photonic interface, which means that it can only address Alice and Bob in a sequential fashion. 
Examples where only one of the qubit memories has an interface to the photonic channel include modular ion traps~\cite{hucul2015modular} and nitrogen-vacancy centres in diamond~\cite{blok2015towards, reiserer2016robust, gao2015coherent}. 
The situation is similar for atoms or ions trapped in a single cavity~\cite{reiserer2015cavity}. In this case, both memories can have a photonic interface. However, typically only one of the interfaces can be active at a given moment.

The figure of merit that we have chosen to evaluate the repeater is the secret-key rate. That is, the ratio between the number of generated secret bits and the number of uses of the quantum channel connecting the two parties. The secret-key rate is a very natural quantifier of the performance of the studied scheme for the task of the secret key generation. It depends both on the success rate of the protocol as well as on the quality of the transmission. We compare the secret-key rate achievable with the repeater with a set of benchmarks that we introduce here. The most strict of these benchmarks is the capacity of the channel~\cite{wilde2013quantum}. That is, the optimal secret-key rate achievable over optical fibre unassisted by a quantum repeater~\cite{pirandola2015fundamental}. The other benchmarks correspond to the optimal rates achievable with additional restrictions. 
In consequence, these benchmarks form a set of stepping stones towards the first quantum repeater able to produce a secure key over large distances.

The idea of assessing quantum repeaters by comparing with the optimal unassisted rates~\cite{takeoka2014fundamental, goodenough2016assessing, pirandola2015fundamental, wilde2016energy, Wilde:2016aa, pirandola2015general, christandl2016relative, bardhan2014strong} has spurred a significant amount of research devoted to developing sophisticated repeater proposals. Analysis of practical systems that utilise only parametric down-conversion sources and optical measurement setups~\cite{khalique2015practical} has shown that such systems do not allow for overcoming the channel capacity, which hints at the importance of quantum memories in repeater architectures. Specific architectures that utilise entangled-photon pair sources together with multimode quantum memories have also been considered in this context~\cite{guha2015rate, krovi_15}. Their analysis suggests that the required efficiency of those entangled-photon pair sources and number of storage modes might be experimentally very challenging for implementation in the very near future. Finally, the so called all-optical repeaters that do not require quantum memories but allow to overcome the channel capacity have been proposed~\cite{pant2017rate}. However, they necessitate the ability to create large photonic cluster states which are beyond current experimental capabilities.

A detailed analysis of a realistic, single-node proof of principle repeater that includes all the specific system imperfections has been recently performed~\cite{luong2015overcoming}. In particular, the analysis identified parameter regimes where it would be possible to surpass the optimal direct transmission rates with a repeater scheme that is close to experimental implementation.
We build upon the analysis of~\cite{luong2015overcoming} by introducing two methods that allow us to achieve higher rates. The first of these methods is the introduction of a maximum storage time for the memories in the quantum repeater. This restriction effectively reduces the effect of decoherence. We derive tight analytical bounds for the secret-key rate as a function of the maximum storage time. In this way we can perform efficient optimisation of the secret-key rate over the maximum storage time. The second of these methods is advantage distillation~\cite{gottesman2003proof}, a two-way classical post-processing technique that allows for distilling secret key at a higher rate than achievable with only one-way post-processing.

The structure of the paper is as follows. In Section~\ref{sec:model} we detail our key distribution protocol. 
The sources of errors, such as losses in the apparatus and noisy operations and storage, are discussed in Section~\ref{sec:sources}. 
In Section~\ref{sec:analysis}, we calculate the secret-key rate that the single sequential quantum repeater would achieve. We define the benchmarks in Section~\ref{sec:assessing}, and in Section \ref{sec:results} we numerically explore the parameter regimes for which the quantum repeater implementation overcomes each benchmark and determine how the secret-key rate of the proposed protocol scales as a function of the distance. We end in Section~\ref{sec:conclusions} with some concluding remarks.

\section{Protocol for a single sequential quantum repeater}
\label{sec:model}
A quantum key distribution protocol consists of two main parts. First, Alice and Bob exchange quantum signals over a quantum channel and measure them to obtain a raw key that is post-processed in a second, purely classical part into a secure key~\cite{scarani2009security}. Here, we focus our interest on the entanglement-based version of the BB84~\cite{Bennett_84} and the six-state~\cite{bruss1998optimal} protocols. In this section, we describe the first part of both key distribution protocols. 

The physical setup consists of two spatially separated parties Alice and Bob connected to an intermediate repeater via optical fibre channels. We note that such a repeater does not need to be positioned exactly half-way between Alice and Bob. The repeater is composed of two qubit quantum memories which we denote by $\textrm{QM}_1$ and $\textrm{QM}_2$. The repeater is then able to generate memory-photon entanglement, where the photonic degree of freedom in which the qubits are encoded is assumed to be time-bin. Alice and Bob each have an optical detector setup that performs a BB84 or a six-state measurement. For technical reasons (see Section~\ref{sec:darkcountsmain}), we consider slightly different setups for BB84 and six-state. More concretely, for BB84 we consider an active setup that switches randomly between the two measurement bases, while in the six-state protocol we consider a passive setup that chooses between the three measurement bases by a passive optical construction~\cite{gittsovich2014squashing}.

Let us now describe a first version of the protocol without a maximum storage time. First, the quantum repeater attempts to generate an entangled qubit-qubit state between a photon and the first quantum memory $\textrm{QM}_1$, after which the photon is sent through a fibre to Alice. Such a \emph{trial} is attempted repeatedly until a photon arrives at Alice's side, after which Alice performs either a BB84 or a six-state measurement.
Second, the quantum repeater attempts to do the same on Bob's side with the second quantum memory $\textrm{QM}_2$ while the state in $\textrm{QM}_1$ is kept stored.
We denote the number of trials performed until a photon arrives at Alice's and Bob's sides $n_A$ and $n_B$ respectively. 
After Bob has received and measured a photon, a Bell state measurement is performed on the two states in $\textrm{QM}_1$ and $\textrm{QM}_2$. We denote by $p_\text{bsm}$ the probability that the measurement succeeds. The classical outcome of the Bell state measurement is communicated to Bob.
This concludes a single \emph{round} of the protocol. We note that in this protocol every round ends with a successful generation of one bit of raw key. 
Such a protocol is closely related to the memory-assisted measurement-device-independent QKD setup (MA-MDI QKD)~\cite{panayi2014memory}. We discuss this connection in Appendix~\ref{sec:MDI}.

One of the main problems in a quantum repeater implementation is that a quantum state will decohere when it is stored in a quantum memory. This means that if it takes Bob a large amount of trials to receive a photon, the state in the quantum memory $\textrm{QM}_1$ will have significantly decohered, preventing the generation of secret key. This motivates the introduction of a \emph{cut-off}. A cut-off is a limit on the amount of trials that Bob can attempt to receive a photon. We denote this maximum number by $\nstar$.

The protocol that we consider here modifies the protocol above as follows: if in a given round Bob reaches the cut-off without success, the round is interrupted and a new round starts from the beginning with the quantum repeater again attempting to send a photon to Alice. In this scheme a large number of rounds might be required until a single bit of raw key is successfully generated.
See Algorithm \ref{algorithm:protocol} for a description of the modified protocol with the cut-off.

{\centering
\begin{minipage}{.85\linewidth}
\begin{algorithm}[H]
\caption{Generation of a bit of raw key with a single sequential quantum repeater }\label{algorithm:protocol}
\begin{algorithmic}[1]
\Initialize{\strut $n_A \gets 0,~n_B \gets 0,~k\gets 0$}
\Loop	
	\State $k\gets k+1$ \Comment{Increment the number of rounds}
	\Repeat	
		\State $n_A\gets n_A + 1$ \Comment{Increment the number of Alice's channel uses}
		\State Generate entangled photon-$\textrm{QM}_1$ pair 
		\State Send entangled photon through fibre towards Alice
	\Until{Alice receives photon}
	\State Alice performs a BB84 or a six-state measurement, stores result
	
	\Repeat 
		\State $n_B\gets n_B + 1$ \Comment{Increment the number of Bob's channel uses}
		\State Generate entangled photon-$\textrm{QM}_2$ pair 
		\State Send entangled photon through fibre towards Bob
	\Until{Bob receives photon or $n_B = k\nstar$}
	\If{Bob received photon}
	 	\State Bob performs a BB84 or a six-state measurement, stores result
		\State Perform the Bell state measurement on the memories, communicate result
		\State Store $\max(n_A,n_B)$ \Comment{Store channel uses}
		\State \Return
	\EndIf
\EndLoop
\end{algorithmic}
\end{algorithm}
\end{minipage}
\par
}

\section{Sources of errors}
\label{sec:sources}
In this section, we model the different elements in the setup to identify the sources of losses and noise. The losses in the system are not only due to the transmissivity of the fibre; depending on the implementation a significant amount of photons is lost before they enter the fibre or due to the non-unit detector efficiency. The causes of noise are the experimental imperfections of the operations, measurements and quantum memories.

\subsection*{Losses}
\begin{center}
\begin{figure}
\centering
\includegraphics[scale=1.05,clip,trim = 45mm 235mm 40mm 20mm]{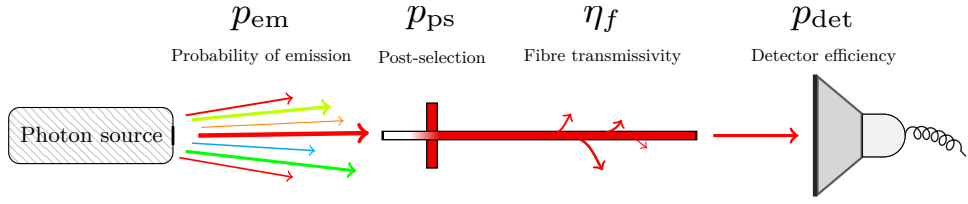}
\caption{General model of all photon losses occurring in the repeater setup. $p_{em}$ is the probability of generating and capturing a photon into the fibre. For experimental reasons a fraction $(1-p_\textrm{ps})$ of photons are additionally filtered out. The fibre has a transmissivity $\eta_{f}$. After exiting the fibre, the photons produce a click in the detector with probability $p_\textrm{det}$. The total efficiency of the apparatus is described by one parameter, $p_{\textrm{app}} = p_{\textrm{em}}p_{\textrm{det}}$.}
\label{fig:losses}
\end{figure}
\end{center}

We model the process of generating and sending an entangled photon through a fibre as follows (see FIG. \ref{fig:losses}). 
First, the photon has to be generated at some photon source and be captured in the fibre. This process happens with probability $p_{\textrm{em}}$. Depending on the experimental implementation, only a fraction $p_{\textrm{ps}}$ of the photons entering the fibre can be used for secret key generation. This can occur for any number of reasons, for instance photons might be filtered according to frequency or a certain time-window~\cite{reiserer2015cavity, gao2015coherent}. The filtering can happen either before or after the transmission through the fibre. The fibre losses are modelled as an exponential decay of the transmissivity $\eta_f$ with the distance $L$, i.e.~$\eta_f = \exp(-\frac{L}{L_0})$ for some fibre attenuation length $L_0$. We denote by $\eta_A$ the fibre losses on Alice's side and by $\eta_B$ the fibre losses on Bob's side. Finally, the arriving photons will be captured by the detectors with an efficiency $p_\textrm{det}$. This probability of detecting a photon will be increased by the presence of dark counts (which will also inevitably add noise to the system), see the discussion of the dark counts at the bottom of this section and in Appendix~\ref{sec:darkcounts}. We define the quantity $p_{\textrm{app}} = p_{\textrm{em}}p_{\textrm{det}}$ describing the total efficiency of our apparatus.

\subsection*{Noise}
\label{sec:noisesection}
We model all noise processes either by the action of a dephasing channel
\begin{equation}
\mathcal D_\text{dephase}^{\lambda_1}(\rho)=\lambda_1\rho+\left(1-\lambda_1\right)Z\rho Z
\end{equation}
or that of a depolarising channel 
\begin{equation}
\mathcal D_\text{depol}^{\lambda_2}(\rho)=\lambda_2\rho+\left(1-\lambda_2\right)\frac{\mathbb{I}}{2}
\end{equation}
where the parameters $\lambda_1$ and $\lambda_2$ quantify the noise, $Z$ is the qubit gate $\left( \begin{smallmatrix} 1&0\\ 0&-1\end{smallmatrix}\right)$ and $\mathbb{I}/2$ is the maximally mixed state. The noise processes occur due to imperfect operations, decoherence of the state while stored in $\textrm{QM}_1$ and dark counts in the detectors.

The noise from imperfect quantum operations is captured by two parameters: $F_{\textrm{prep}}$ and $F_{\textrm{gm}}$. $F_{\textrm{prep}}$ is a dephasing parameter which corresponds to the preparation fidelity of the memory-photon entangled state~\cite{togan2010quantum}. $F_{\textrm{gm}}$ is a depolarising parameter that describes the noise introduced by the imperfect gates and measurements performed on the two quantum memories during the protocol \cite{cramer2015repeated, kalb2017entanglement}. Hence, the noise can be modelled by a dephasing and a depolarising channel with $\lambda_1=F_{\textrm{prep}}$ and $\lambda_2=F_{\textrm{gm}}$.

The decoherence is modelled by a decay of the fidelity in the number of trials $n$. This decoherence is caused by two distinct effects. Firstly, there is the decoherence due to the time that the quantum repeater has to wait between sending photons. This time is the time it takes to confirm whether the photon got lost plus the time it takes to generate a photon entangled with the memory. We model this effect through an exponential decay of fidelity with time \cite{nemoto2016photonic}, which is expected whenever excess dephasing is suppressed (e.g.~by dynamical decoupling~\cite{de2010universal}). However, we note that this is not the only possible model of decay, in several experiments a Gaussian decay has been observed~\cite{specht2011single, hucul2015modular,sangouard2011quantum, thiel2011rare}. Secondly, attempting to generate an entangled photon-memory pair at $\textrm{QM}_2$ might also decohere the state stored in the $\textrm{QM}_1$. For example, this effect is the most prominent decoherence mechanism in nitrogen-vacancy implementations~\cite{reiserer2016robust}, where an exponential decay of fidelity with the number of trials was observed. This is also how we model that effect here.

The quantum state $\rho$ that is subjected to those effects undergoes an evolution given by the dephasing and depolarising channels 
with $\lambda_1=(1+e^{-an})/2$ and $ \lambda_2= e^{-bn}$. The two parameters $a$ and $b$ are given by
\begin{gather}
a = a_0+a_1\left(\frac{2n_{\textrm{ri}}L_B}{c}+t_{\textrm{prep}}\right)\ ,\\
b = b_0+b_1\left(\frac{2n_{\textrm{ri}}L_B}{c}+t_{\textrm{prep}}\right)\ ,
\end{gather}
where $n_{\textrm{ri}}$ is the refractive index of the fibre, $c$ is the speed of light in vacuum, $L_B$ the distance from the quantum repeater to Bob and $t_{\textrm{prep}}$ is the time it takes to prepare for the emission of an entangled photon. Here $a_0$ and $b_0$ quantify the noise due to a single attempt at generating an entangled state and $a_1$ and $b_1$ quantify the noise during storage per second. 
\label{sec:darkcountsmain}
Finally, the dark counts in the detectors introduce depolarising noise. This model is justified for the two quantum key distribution protocols that we consider, see~\cite{gittsovich2014squashing,beaudry2008squashing}.
We let $\alpha_{A/B}$ denote the corresponding depolarising parameter on Alice's/Bob's side. The details of this model are presented in Appendix~\ref{sec:darkcounts}.

\section{Secret-key rate of a single sequential quantum repeater }
\label{sec:analysis}
The secret-key rate $R$ is defined as the amount of secret-key bits generated by a protocol divided by the number of channel uses and the number of optical modes.
In the particular case of our sequential quantum repeater, the secret-key rate is given by 
\begin{equation}
R = \frac{Y}{2}r\ .
\end{equation} 
The yield $Y$ of the protocol is defined as the rate of raw bits per channel use. The secret-key fraction $r$ is defined as the average amount of secret key that can be extracted from a single raw bit. The factor of a half is due to the fact that the encoding uses two optical modes. 
Since we consider two possible quantum key distribution protocols we take 
\begin{equation}
r = \max\{r_{\text{BB84}}, r_{\text{six-state}}\} \, .
\end{equation}
where $r_{\text{BB84}}$ and $r_{\text{six-state}}$ are the secret-key fractions of the BB84 and six-state protocols, respectively (see Eq.~\eqref{eq:bb84skf} and Appendix \ref{sec:secretkeyfracderiv}).

\subsection*{Yield}
The yield can be calculated as $p_{\textrm{bsm}}$ (i.e.~the success probability of the Bell state measurement) divided by the (average) 
number of channel uses needed for the successful detection of a photon by both Alice and Bob in the same round. With a single sequential quantum repeater it is not obvious how to count the number of channel uses. As in~\cite{luong2015overcoming}, we count the \emph{maximum} of the two channel uses on Alice's and Bob's sides respectively,

\begin{align}
Y = \frac{p_{\textrm{bsm}}}{\mathbb{E}\left[N\right]} = \frac{p_{\textrm{bsm}}}{\mathbb{E}\left[\max(N_A, N_B)\right]}\label{eq:yielddefinition}\ .
\end{align}
where $N$, $N_A$ and $N_B$ are the random variables that model the number of channel uses, the number of channel uses at Alice's side and the number of channel uses at Bob's side, respectively.

Without the cut-off, it is possible to obtain an analytical formula for the average number of channel uses~\cite{panayi2014memory, luong2015overcoming},
\begin{equation}
\mathbb{E}\left[\max(N_A, N_B)\right] =
       \frac{1}{p_A}+\frac{1}{p_B}-\frac{1}{p_A+p_B-p_Ap_B}\ ,
\end{equation} 
where $p_A$ and $p_B$ depend on the quantum key distribution protocol and are given by the following equations (see Appendix \ref{sec:darkcounts}),
\begin{align}
p_{A/B,\textrm{BB84}} = 1 - (1-p_{\textrm{app}} p_{\textrm{ps}}\eta_{A/B})(1-p_d)^2\ , \label{eq:ps1} \\
p_{A/B,\textrm{six-state}} = 1 - (1-p_{\textrm{app}} p_{\textrm{ps}}\eta_{A/B})(1-p_d)^6\ .
\label{eq:ps2}
\end{align}
Here $p_d$ is the probability of measuring a dark count.

Every time that Bob reaches $\nstar$ trials, Alice and Bob restart the round and start over again. The cut-off thus increases the average number of channel uses. We have developed an analytic approximation of $\mathbb{E}\left[N\right]$ which is essentially tight (see Appendix \ref{sec:yieldderiv} for the derivation and error bounds)
\begin{align}
\mathbb{E}\left[\max(N_A, N_B)\right] \approx  \left\{
\begin{array}{ll}
      \frac{1}{p_A\left(1-\left(1-p_B\right)^{\nstar}\right)} & \frac{1}{p_A}> \nstar \\
       \frac{1}{p_A}+\frac{1}{p_B}-\frac{1}{p_A+p_B-p_Ap_B} & \frac{1}{p_A}\leq\nstar\ . \\
\end{array} 
\right.
\end{align}

\subsection*{Secret-key fraction}
Here we consider the secret-key fraction of the BB84 and six-state protocols. As we discussed previously, we consider the BB84 protocol with an active measuring scheme and the six-state protocol with a passive one. 
Moreover, we consider a fully asymmetric version of BB84 and a fully symmetric version of six-state. Fully symmetric means that all bases are used with equal probability while fully asymmetric means that the ratio at which one of the bases is used is arbitrarily close to one. Finally, we consider a one-way key distillation scheme for BB84~\cite{scarani2009security} while for the six-state protocol we consider the advantage distillation scheme in~\cite{watanabe2007key}. Advantage distillation~\cite{gottesman2003proof} is a classical post-processing technique that allows to increase the secret-key fraction at all levels of noise.

The reasons for not analysing the BB84 protocol with advantage distillation and the fully asymmetric six-state with advantage distillation are technical. In the case of BB84, computing the rate with advantage distillation requires the optimisation over a free parameter. The combination of the optimisation over the cut-off together with the extra free parameter was computationally too intensive to consider here.

For the six-state protocol there is, to our knowledge, no security proof that can deal with the asymmetric six-state protocol with photonic qubits without introducing extra noise~\cite{gittsovich2014squashing, ballester2008state}. However, these protocol choices do not have a strong impact on our analysis. Advantage distillation does not significantly increase the amount of distillable key for low error rates. Hence, asymmetric BB84 without advantage distillation is only slightly suboptimal. For higher error rates, where advantage distillation plays a role, the symmetric six-state protocol with advantage distillation is a factor of three away from the asymmetric version.

The expression for the secret-key fraction of both protocols depends on the error rates in the $X$, $Y$ and $Z$ bases, which we denote by $e_X$, $e_Y$ and $e_Z$. In the case of the BB84 protocol,~\cite{scarani2009security, lo2005efficient} it is given by
\begin{equation}
r_{\text{BB84}} = 1 - h(e_Z) - h(e_{X})\ ,\label{eq:bb84skf}
\end{equation} 
where $h(p) = - p\log_2 p - (1-p) \log_2 (1-p)$ is the binary entropy function. The expression for $r_{\text{six-state}}$ is more complex; we leave its discussion to Appendix~\ref{sec:secretkeyfracderiv}. 

We can directly evaluate the error rates in each basis as a function of the general parameters of Section \ref{sec:noisesection}. For the single sequential quantum repeater these average errors are
\begin{equation}
e_X = e_Y = e_{XY} =\frac{1}{2}-\frac{1}{2}F_{\textrm{gm}}\alpha_A \alpha_B\left(2F_{\textrm{prep}}-1\right)^2\left\langle e^{-\left(a+b\right)n} \right\rangle\ ,\label{eq:avgex}\\
e_Z = \frac{1}{2}-\frac{1}{2}F_{\textrm{gm}}\alpha_A \alpha_B\langle e^{-bn}\rangle\label{eq:avgez}\ .
\end{equation}
where $\left\langle e^{-cn}\right\rangle$ is the average of the exponential $e^{-cn}$ over a geometric distribution over the first $\nstar$ trials. 
The detailed derivation of the error expressions is presented in Appendix~\ref{sec:QBER}.

\section{Benchmarks for the assessment of quantum repeaters}
\label{sec:assessing}
We introduce a set of benchmarks to assess the performance of a quantum repeater implementation. 

The first benchmark that we consider is the rate that would be achieved with the same parameters for the system losses and dark counts and for the same protocol but without a quantum repeater. Overcoming this benchmark gives the first indication that the repeater setup is useful; it means that the repeater setup outperforms the setup without repeater. We call this benchmark the direct transmission benchmark.

The remaining benchmarks represent the optimal secret-key rate that Alice and Bob could achieve if they were to communicate over the same quantum channel without a repeater under some constraints.

The optimal secret-key rate without a repeater highly depends on the channel model. The first modelling decision is the placement of the boundary between Alice's and Bob's laboratories and the quantum channel.
This is because it is not \emph{a priori} clear where the channel begins and ends. However, this decision has a strong impact on the optimal achievable rate; if the channel includes most of Alice's and Bob's laboratories, then the channel is more lossy and noisy and the benchmark is easier to overcome. If, on the other hand, the channel is just the optical fibre cable the benchmark becomes more difficult to overcome.\\

We consider three cases in terms of the individual lossy components of our setup (see FIG.~\ref{fig:modelofsetup}, FIG.~\ref{fig:losses} and their captions):
\itemsep-0.9mm 
\itemindent-2cm
\begin{itemize}
\setlength{\itemindent}{1cm}
\item[Case 1:] Fibre only, in this case the transmissivity is: $\eta = \eta_{\textrm{f}}=\eta_A\eta_B$.
\item[Case 2:] Fibre and different filters, then the channel transmissivity becomes: $\eta = \eta_{\textrm{f}}p_{\textrm{ps}}$.
\item[Case 3:] Fibre, filters and Alice's and Bob's apparatus, then the transmissivity becomes: $\eta = \eta_{\textrm{f}}p_{\textrm{ps}}p_{\textrm{app}}$.
\end{itemize}
Note that although in the experimental implementation of the repeater the terms $p_{\textrm{ps}}$ and $p_{\textrm{app}}$ appear twice in the expression of the transmissivity, they appear only once in the benchmarks which include them. The reason is that in a scenario without a repeater the emission inefficiency and the filters only affect the transmissivity once. 

The second design parameter for these benchmarks is the type of channel.
Transmission of photons through fibres is modelled as a pure-loss channel~\cite{Weedbrook:2012aa}, where only a fraction $\eta$ of the input photons reach the end of the channel. The first type of channel that we consider is the pure-loss channel without any additional restriction. The optimal achievable rate over one mode of the pure-loss channel is given by the secret-key capacity~\cite{pirandola2015fundamental}
\begin{align}
-\log_2\left(1-\eta\right)\label{eq:capacity}\ .
\end{align}
Note that for high losses the scaling of this capacity with distance is proportional to $\eta_f = \exp\left(-\frac{L}{L_0}\right)$. At the same time with an ideal (noiseless) single quantum repeater placed half-way between Alice and Bob, the expected secret-key rate would scale proportionally to $\sqrt{\eta_f} = \exp(-\frac{L}{2L_0})$~\cite{luong2015overcoming}.

The second type of channel that we consider is the pure-loss channel when the transmitter has a limitation in the energy that can be introduced into the channel. There has been some recent work studying the optimal rate per mode of the finite-energy pure-loss channel~\cite{takeoka2014squashed,goodenough2016assessing,wilde2016energy}. However, the optimal rate remains unknown. The bound that we consider here~\cite{takeoka2014squashed} is given by
\begin{align}
g\left(\left(1+\eta\right)P/2\right) - g\left(\left(1-\eta\right)P/2\right)\label{eq:finiten}\ ,
\end{align}
where $g(x) := \left(x+1\right)\log_2(x+1)-x\log_2x$ and $P$ is the mean photon number. In our repeater setup, the finite energy restriction arises from the fact that, on average, only a fraction of a photon enters the fibre in each trial. More precisely, the average photon number satisfies $P = p_{\textrm{em}}$ in cases 1 and 2 above and $P=1$ in case 3. Unfortunately, since Eq.~\eqref{eq:finiten} is an upper bound, it is only strictly smaller than the capacity of the pure-loss channel for small mean photon number.
Expanding the bounds from equations Eq.~\eqref{eq:capacity} and Eq.~\eqref{eq:finiten} around $\eta = 0$ shows that the cross-over between the two bounds occurs when $p_{\textrm{em}}\log_2\left(\frac{p_{\textrm{em}}+2}{p_{\textrm{em}}}\right) = \frac{1}{\ln2}$. In other words, for high losses the finite-energy bound is tighter when $p_{\textrm{em}}\lesssim 0.796$. This implies that the finite-energy bound does not yield an interesting benchmark in case 3.

The third type of channel that we consider is the thermal-loss channel. An upper bound on the capacity of the thermal-loss channel is
\begin{gather}
-\log_2[\left(1-\eta\right)\eta^{\overline{n}}]-g\left(\overline{n}\right),
\end{gather}
if $\overline{n}<\frac{\eta}{1-\eta}$ and zero otherwise~\cite{pirandola2015fundamental}. Here, $\overline{n}$ is the average number of thermal photons per channel use~\cite{Weedbrook:2012aa}. This is an interesting channel because the effect of dark counts can be seen as caused by the thermal photons. Hence this type of channel becomes relevant for case 3, where detectors, and therefore also the dark counts, are regarded as part of the channel. The details of the dark count model are presented in Appendix~\ref{sec:darkcounts}. There we also show how to easily convert the experimentally relevant dark count rate of the detector and the duration of the detection window $t_{\textrm{int}}$ into $\overline{n}$ and $p_d$, the probability of getting a dark count within the given time window.

The combinations of a channel boundary together with a channel type give us a set of benchmarks. Not all combinations yield interesting benchmarks. In Table~\ref{table:table1}, we summarise the benchmarks that we consider.\\

\begin{table}[H]
\begin{center}
\begin{tabular}{ c|c|c|c|c } 
 & Infinite & Finite & Thermal & Direct transmission \\ 
 \hline
Case 1: $\eta_{\textrm{f}}$ & 1a & 1b & $-$ & $-$\\ 
Case 2: $\eta_{\textrm{f}}p_{\textrm{ps}}$ & 2a & 2b & $-$ & $-$\\ 
Case 3: $\eta_{\textrm{f}}p_{\textrm{ps}}p_{\textrm{app}}$ & $-$ & $-$ & 3c & 3d \\ 
\end{tabular}
\caption{Labels of the benchmarks that we use to assess the performance of a quantum repeater. These labels are frequently referred to in the numerical results. Each row corresponds to a different channel boundary, which translates into an effective channel transmissivity. Each column corresponds to a different type of channel: pure loss, pure loss with energy constraint and thermal channel, and the final column corresponds to the direct transmission benchmark. }
\label{table:table1}
\end{center}
\vspace*{-10mm}
\end{table}

\section{Implementation based on Nitrogen-Vacancy centre setup}
\label{sec:NV}

Our model is fully general and can be applied to a wide range of physical platforms. To illustrate its performance we will now consider one of such potential near-term realisations of a single sequential quantum repeater. For this particular example we choose to base our system on Nitrogen-Vacancy (NV) centres in diamond. NVs are a prime candidate for this task due to their optical interface featuring high-fidelity single-shot readout~\cite{robledo2011high} and their recently demonstrated capabilities to distribute spin-photon entanglement while faithfully storing quantum states~\cite{kalb2017entanglement}.

\begin{figure}
	\centering
	\includegraphics{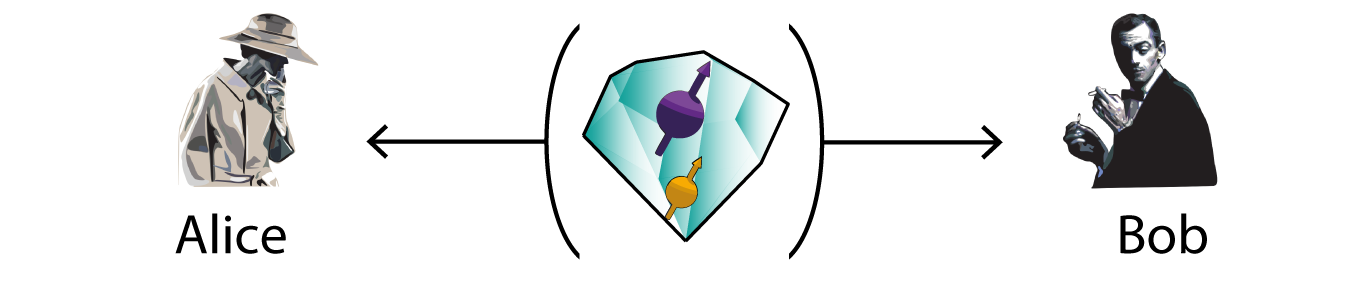}
	\caption{\label{fig:ExpSetup} Single sequential quantum repeater based on an electron spin associated with an NV (purple) and $^{13}$C nuclear spin (orange) in diamond. The previous quantum memories $\mathrm{QM}_{1,2}$ are now represented by the electron and nuclear spin respectively. The optical interface of the NV is strongly Purcell-enhanced by an optical cavity with low-mode volume and allows for efficient photon transmission to Alice and Bob.}
\end{figure}

In the following we expand on the required experimental techniques (see Fig.~\ref{fig:ExpSetup}). The NV centre itself can be readily used as a generator of spin-photon entanglement at cryogenic temperatures. The NV is encapsulated in an optical cavity of low-mode volume~\cite{riedel2017deterministic} to strongly enhance the emission into the zero phonon line (ZPL) via the Purcell effect. As no particular low-loss cavity design has been implemented with NVs yet, we rely purely on the aforementioned ZPL enhancement. More specific cavity configurations that allow for reflection based mechanisms rely on the realisation of a low-loss overcoupled cavity to be efficient \cite{duan2004scalable} and might become available in the future.

Firstly, we generate spin-photon entanglement~\cite{pfaff2014unconditional} and send the emitted photon off to Alice who reports successful detection events back to the repeater station. Note that electron spin decoherence during communication rounds is negligible since second-long coherence times have been demonstrated by employing XY8 dynamical decoupling sequences~\cite{abobeih2018one}.

Upon success the optical interface of the NV is reused for communication with Bob. To this end, the NV spin state that is correlated with Alice's measurement outcome is stored on a $^{13}$C nuclear spin in the vicinity of the electron spin, which itself is then reinitialised. We choose a configuration in which the always-on magnetic hyperfine coupling between both spins is weak (on the order of a few $\mathrm{kHz}$). This configuration has been experimentally shown to result in a highly-addressable quantum memory which is resilient to optical excitation and reinitialisation of the NV spin~\cite{reiserer2016robust}. Coherently swapping the NV state onto - and high-fidelity control over - such a weakly-coupled nuclear spin has been demonstrated recently~\cite{taminiau2014universal,kalb2017entanglement}.

The protocol then proceeds as described in Section~\ref{sec:model} by communicating with Bob. Note that repeated communication attempts will eventually decohere the memory state due to the necessity for frequent electron spin resets and the always-on hyperfine interaction between the two spins. This constitutes the main source of error in this system (parametrised by $a_0$ and $b_0$, see Sec.~\ref{sec:noisesection}).

After a successful state transmission to Bob, we conduct a sequential two-step Bell state measurement and read-out the $XX$ and $ZZ$ parities of the combined nuclear-electron spin state, where $X$ and $Z$ denote the standard Pauli matrices. This can be achieved by means of the earlier mentioned universal control over the system or by introducing additional resource qubits such as the nitrogen nuclear spin associated with the NV~\cite{pfaff2014unconditional}.

\section{Numerical results}
\label{sec:results}
In this section, we perform a numerical analysis of our model applied to the physical system based on NV centres as described in Section~\ref{sec:NV}. All numerical results have been obtained using a Mathematica notebook~\cite{Note1}.
Unless specified otherwise, we use the following parameters that we call ``expected parameters''. These parameters represent best-case scenarios from the chosen references. These experimental capabilities do not fundamentally contradict or exclude each other and seem therefore achievable in a single experimental NV setup. 
\begin{itemize}
\item $a_0 \, \textrm{(dephasing due to interaction)}  = \frac{1}{2000}$ per attempt~\cite{reiserer2016robust},
\item $a_1 \, \textrm{(dephasing with time)} = \frac{1}{3}$ per second~\cite{maurer2012room},
\item $b_0 \, \textrm{(depolarisation due to interaction)} = \frac{1}{5000}$ per attempt~\cite{reiserer2016robust},
\item $b_1 \, \textrm{(depolarisation with time)} = \frac{1}{3}$ per second~\cite{maurer2012room},
\item $t_{\textrm{prep}} \, \textrm{(memory-photon entanglement preparation time)} = 6 \, \mu  \textrm{s}$~\cite{hensen2015loophole},
\item $F_{\textrm{gm}} \, \textrm{(depolarising parameter for gates and measurements)} = 0.9$~\cite{kalb2017entanglement},
\item $F_{\textrm{prep}} \, \textrm{(dephasing parameter for the memory-photon state preparation)} = 0.99$ ~\cite{hensen2015loophole},
\item $p_{\textrm{em}} \, \textrm{(probability of emission)} = 0.49$~\cite{hensen2015loophole, bogdanovic2016design},
\item $p_{\textrm{ps}} \, \textrm{(post-selection)} = 0.46$~\cite{riedel2017deterministic},
\item $p_{\textrm{det}} \, \textrm{(detector efficiency)} = 0.8$~\cite{hensen2015loophole},
\item $p_{\textrm{bsm}} \, \textrm{(Bell state measurement success probability)} =1$~\cite{pfaff2014unconditional},
\item $\textrm{Dark count rate} = 10 \, \, \textrm{per second}$~\cite{hensen2015loophole},
\item $t_{\textrm{int}}\, (\textrm{detection window}) = 30$ ns~\cite{hensen2015loophole},
\item $L_0 \, \textrm{(attenuation length)}= 0.542$ km~\cite{hensen2015loophole},
\item $n_{\textrm{ri}} \, \textrm{(refractive index of the fibre)}= 1.44$~\cite{LaserEncyclopedia}.
\end{itemize}

Before we present the results, we note that the emission frequency of the nitrogen-vacancy centres results in a relatively low $L_0$ which in turn does not allow to achieve large distances.
In practical quantum key distribution networks, assuming that dedicated fibres are used for which one can choose which frequency mode one wants to transmit at, this problem might be overcome using the frequency conversion of the emitted photons into a telecom frequency, which will yield an increased $L_0$. Note that the benchmarks in Table \ref{table:table1} will scale accordingly. There is a range of frequencies used in fibre-based communication and for each of those frequencies the attenuation length varies greatly depending on the type of the fibre used. To give some examples, the best fibres at $1560$ nm have losses of $0.1419$ dB/km ($L_0 \approx 30.6$ km) \cite{tamura2017lowest}, while at $1310$ nm standard single-mode fibres exhibit losses of $0.4$ dB/km ($L_0 \approx 10.9$ km)~\cite{keiser2011optical}. 
Clearly our model is general and can be applied to a channel with any value of $L_0$. Here, throughout most of this section, we consider the transmission through the channel at the same wavelength as the emission line of the NV-centre setup, as such a channel for this specific physical system has been realised in an experiment~\cite{hensen2015loophole} using fibre with losses of $8$ dB/km ($L_0 = 0.542$ km as given in the list of parameters above). At the end we present an additional plot describing the scenario in which a telecom channel with the commonly used in the quantum repeater community attenuation length of $L_0 \approx 22$ km is available. In this case the frequency conversion of the emitted photons to telecom is applied.

\begin{figure}
\centering
\includegraphics[scale=0.8,clip,trim = 0mm 25mm 0mm 20mm]{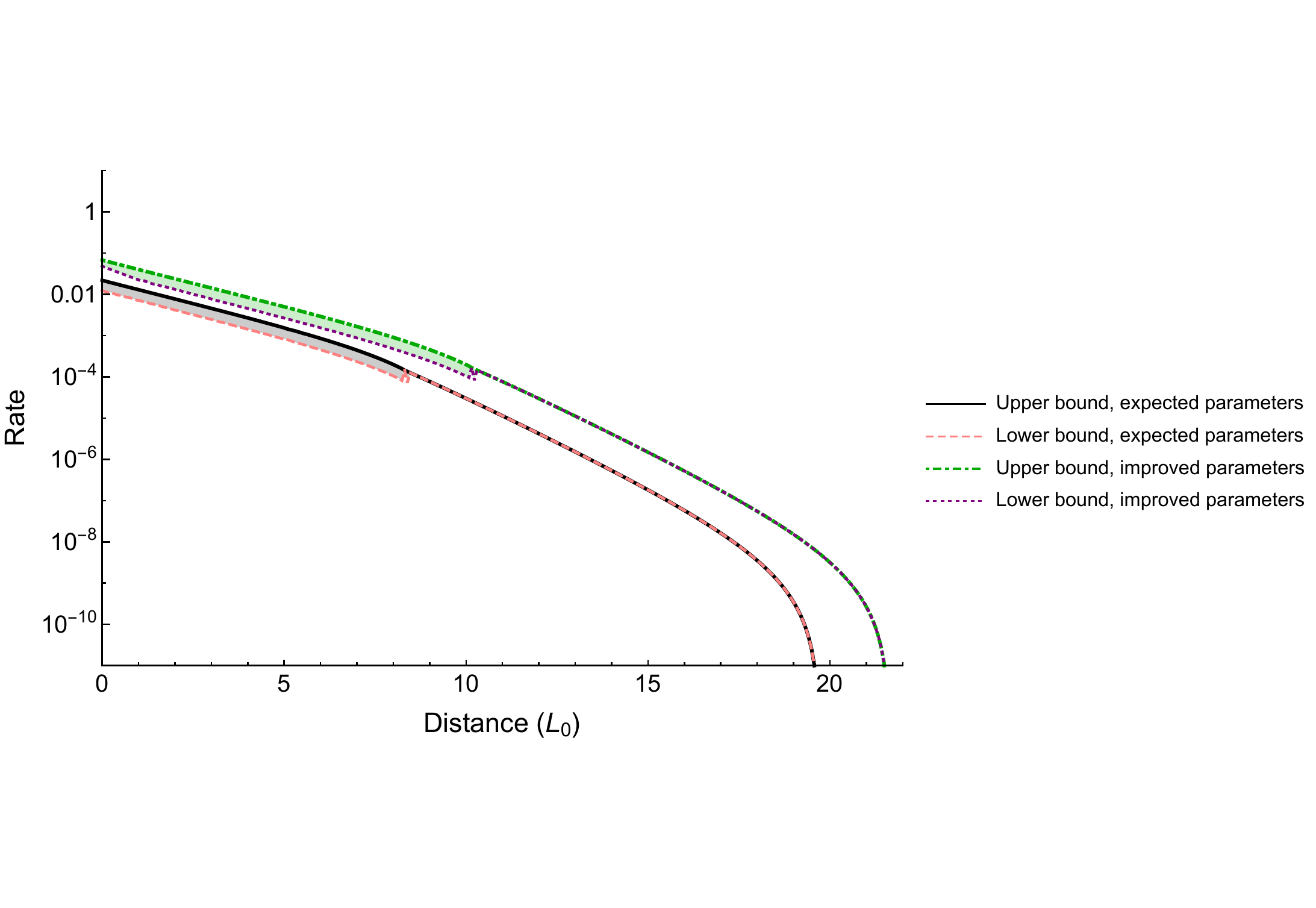}
\caption{Upper- and lower bounds on the secret-key rate rate with a quantum repeater as a function of the distance in units of $L_0 = 0.542 \text{ km}$. The repeater is positioned half-way between Alice and Bob. The curves correspond to the expected and improved parameters with optimised cut-off. The improved parameters correspond to setting $p_{\textrm{ps}} = p_{\textrm{em}} = 0.6$ and $F_{\textrm{gm}} = 0.97$. For high losses, the upper- and lower bounds become essentially tight. For this reason, the upper bound on the achieved rate forms a reliable estimate of the secret-key rate.}
\label{fig:distanceplot1}
\end{figure}

\smallskip
\textit{Tightness of the error bounds for the secret-key rate.} We have derived upper and lower bounds on the yield, and thus also on the secret-key rate, for the two studied protocols. In FIG. \ref{fig:distanceplot1}, we plot both the upper and the lower bound on the achieved rate with the current and improved parameters ($p_{\textrm{ps}} = p_{\textrm{em}} = 0.6$ and $F_{\textrm{gm}} = 0.97$) and optimised cut-off as a function of the distance in units of $L_0$. 
There are two regimes visible on the plot. This is a consequence of the fact that our bounds have a different analytical form in the two regimes (see Appendix \ref{sec:yieldderiv}). Since for practical purposes our bounds are essentially tight, from now on we will refer to the upper bound as the expected secret-key rate, and will omit the lower bound for the legibility of the plots. 

\begin{figure}[h]
\centering
\includegraphics[scale=0.8,clip,trim = 0mm 25mm 0mm 20mm]{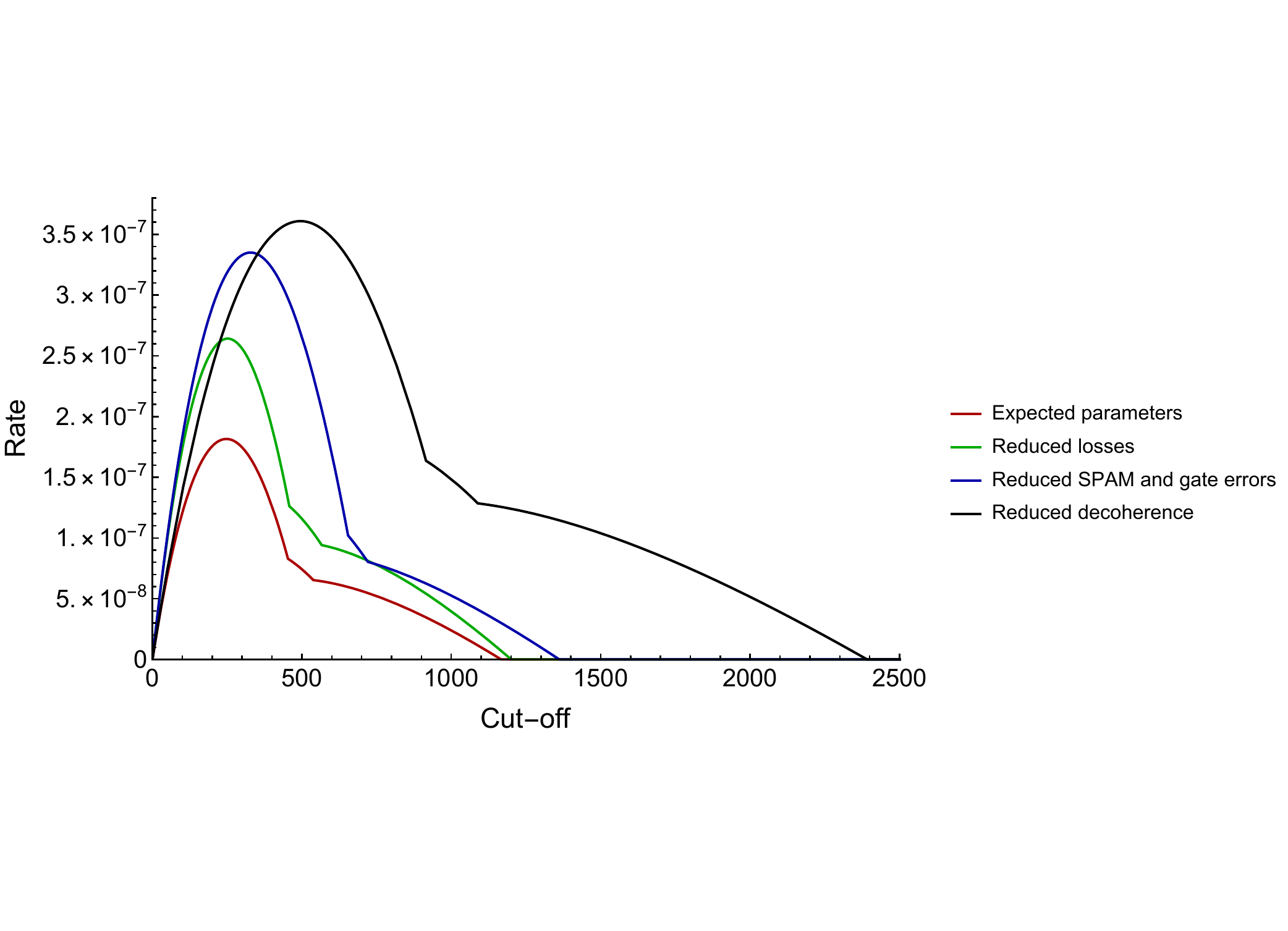}
\caption{Secret-key rate as a function of the cut-off for the expected parameters with the repeater positioned half-way between Alice and Bob. The reduced losses are for $p_{\textrm{app}}' = (p_{\textrm{app}})^{0.9}$ and $p_{\textrm{ps}}' = (p_{\textrm{ps}})^{0.9}$, the reduced SPAM (state preparation and measurement) and gate errors are for $F_{\textrm{gm}}' = (F_{\textrm{gm}})^{0.7}$ and $F_{\textrm{prep}}' = (F_{\textrm{prep}})^{0.7}$ and the reduced decoherence is for $a' = a/2$ and $b'=b/2$. The optimal $\nstar$ shifts depending on the parameters. The kinks arise due to the fact that we optimise over two protocols: fully asymmetric BB84 and symmetric six-state protocol with advantage distillation which itself consists of two subprotocols. The optimal protocol depends on the bit error rates. The data have been plotted for the distance of $15L_0$, where $L_0= 0.542 \text{ km}$.
}
\label{fig:plot1}
\end{figure}

\smallskip
\textit{The impact of the cut-off on the secret-key rate.} In FIG.~\ref{fig:plot1} we plot the secret-key rate versus the cut-off for different sets of parameters. The repeater is assumed to be positioned half-way between Alice and Bob. We observe a strong dependency of the secret-key rate on the cut-off. In particular, for large cut-off the secret-key rate drops to zero. 
This is due to the inclusion of rounds where the state has significantly decohered. This implies that the cut-off is essential for generating a key at large distances. Moreover, we observe that the optimal cut-off highly depends on the explored parameter regime. 

\begin{figure}
\centerfloat
\includegraphics[scale=0.8,clip,trim = 0mm 25mm 0mm 20mm]{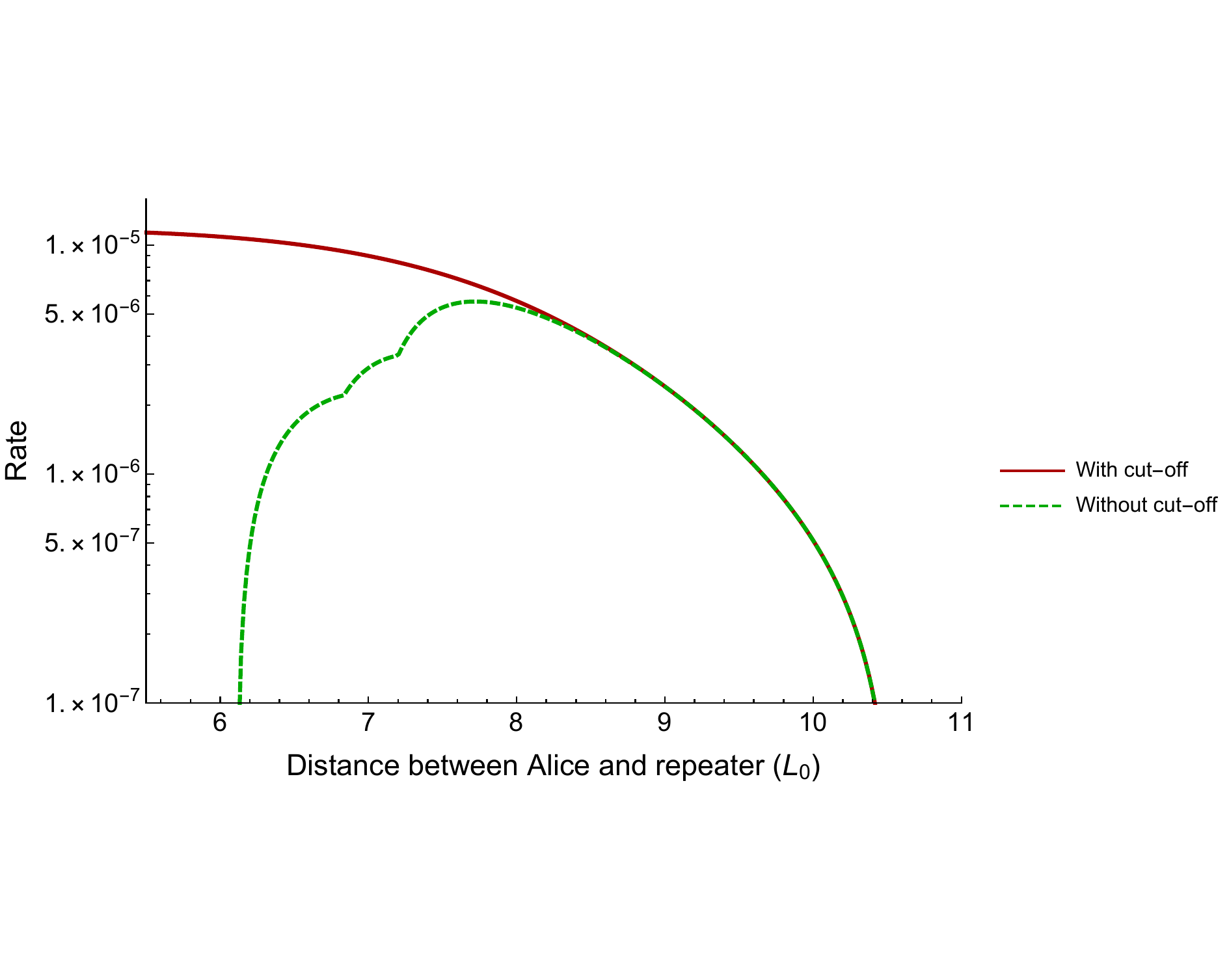}
\caption{Secret-key rate with and without the cut-off as a function of the distance in units of $L_0= 0.542 \text{ km}$ between Alice and quantum repeater. The total distance between Alice and Bob is fixed to $11L_0$. We see that with the cut-off optimisation, positioning the repeater half-way between Alice and Bob is optimal. This behaviour was also observed for other parameter regimes. This result contrasts with the optimal positioning for the no cut-off scenario, for which we see that shifting the repeater towards Bob is beneficial. We also note that the two rates overlap when the repeater is shifted towards Bob.}
\label{fig:optposition}
\end{figure}

\smallskip
\textit{Optimal positioning of the repeater.} The asymmetry of the studied sequential protocol raises the question of whether it is best to position the repeater half-way between Alice and Bob. In fact, in the absence of a cut-off this is not the case~\cite{luong2015overcoming}. For sufficiently large distances, shifting the repeater towards Bob can increase both the secret-key rate and the distance over which the secret-key rate is non-zero in the presence of dark counts. Specifically, 
the optimal positioning remains a fixed distance away from Bob independently of the actual total distance. 
Here, we find that with the cut-off and for the parameters considered this phenomenon disappears. We see in FIG.~\ref{fig:optposition} that the optimal position with the cut-off optimisation appears to be exactly in the middle of Alice and Bob. Nevertheless, we note that the bounds for the yield derived in Appendix~\ref{sec:yieldderiv} are valid under the condition $\eta_B \ge \eta_A$. This means that we can only study the effect of moving the repeater towards Bob. However, we do not expect any benefit in shifting the repeater towards Alice as this could only increase the noise due to decoherence. From now on for the scenarios with the cut-off optimisation, we always consider the repeater to be placed half-way between Alice and Bob. Interestingly, in FIG.~\ref{fig:optposition} we also see that the rates for the two scenarios with and without the cut-off start to coincide after the quantum repeater is shifted within a certain distance of Bob. Intuitively this happens when the probability of Bob getting a photon is large enough so that the significance of the cut-off becomes marginal.

\begin{figure}
\centering
\includegraphics[scale=0.8,clip,trim = 0mm 25mm 0mm 20mm]{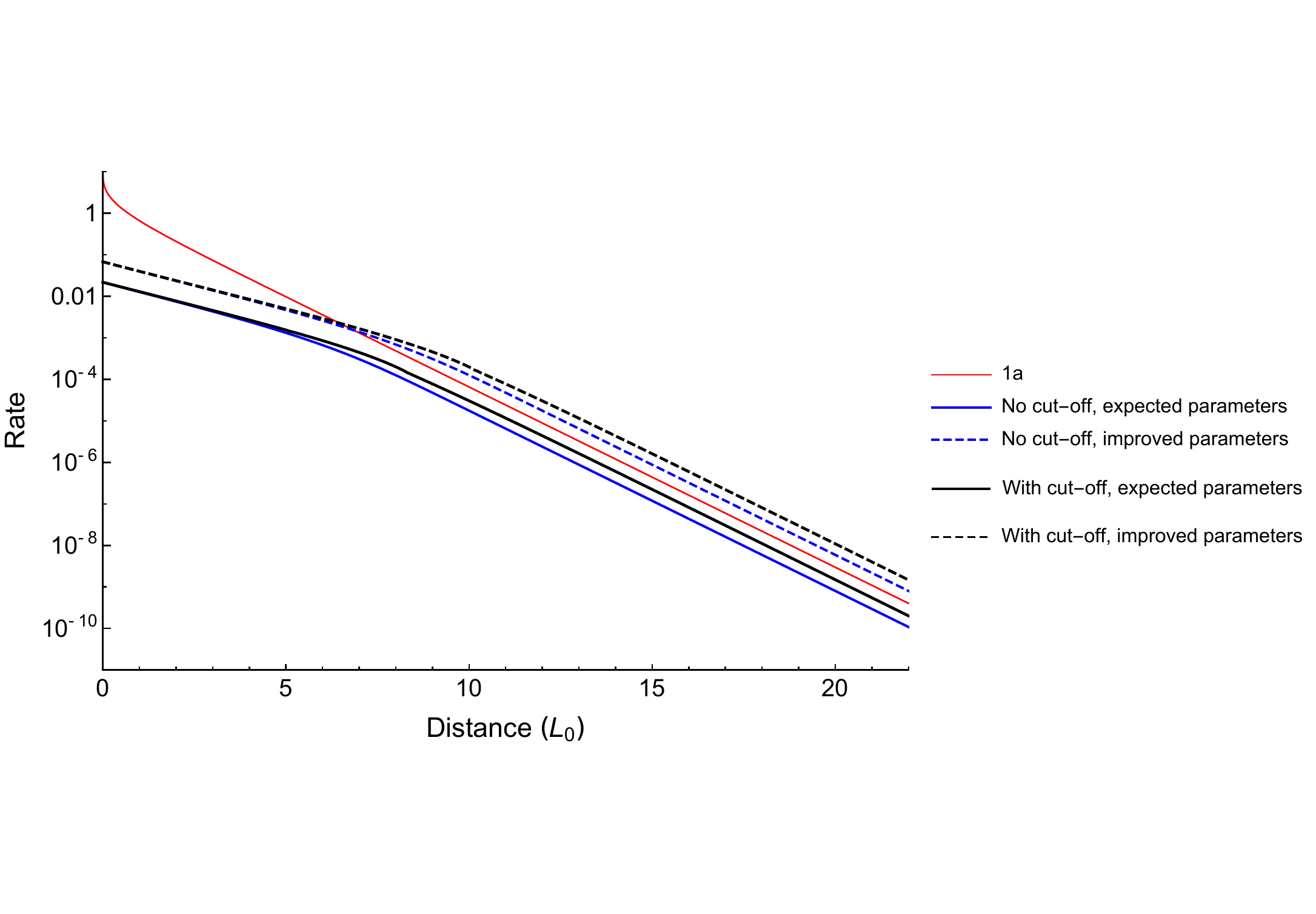}
\caption{Secret-key rate as a function of the distance in units of $L_0= 0.542 \text{ km}$, assuming detectors without dark counts. The black lines correspond to the protocol with cut-off and the blue lines to the protocol without the cut-off but with optimised positioning of the repeater. We plot the data for both the expected and improved parameters. The improved parameters correspond to setting $p_{\textrm{ps}} = p_{\textrm{em}} = 0.6$ and $F_{\textrm{gm}} = 0.97$. Finally, the channel capacity (1a) is also included for comparison. It can be seen that both the cut-off and repositioning of the repeater allows to generate key for all distances.}
\label{fig:distanceplotNoDarkCounts}
\end{figure}

\begin{figure}
\centering
\includegraphics[scale=0.8,clip,trim = 0mm 25mm 0mm 20mm]{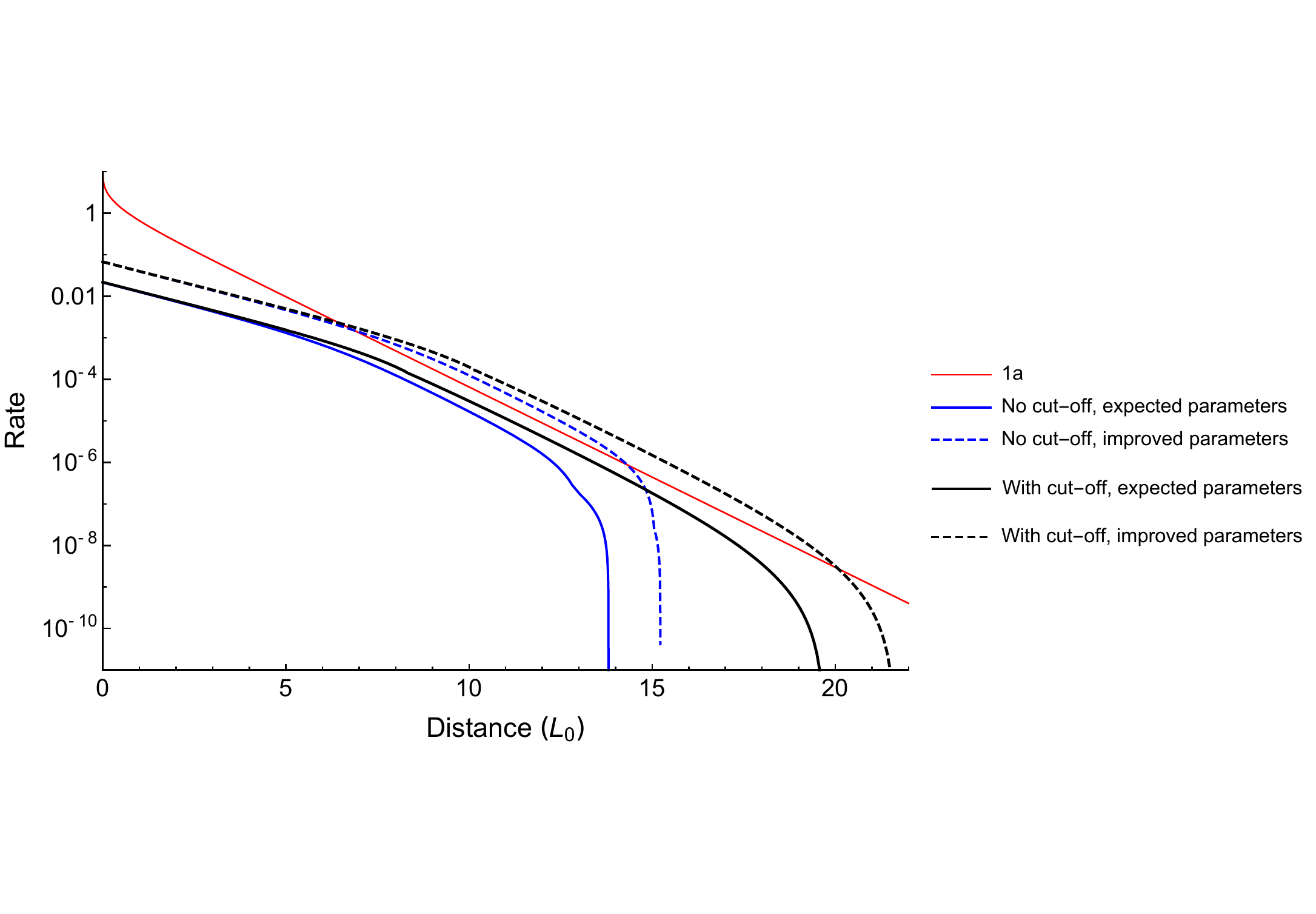}
\caption{Secret-key rate as a function of the distance in units of $L_0= 0.542 \text{ km}$ with dark counts. The black lines correspond to the protocol with cut-off and the blue lines to the protocol without the cut-off but with optimised positioning of the repeater. We plot the data for both the expected and improved parameters. The improved parameters correspond to setting $p_{\textrm{ps}} = p_{\textrm{em}} = 0.6$ and $F_{\textrm{gm}} = 0.97$. Finally, the channel capacity (1a) is also included for comparison. It can be seen that the protocol with the cut-off is more robust against dark counts than the protocol without the cut-off.}
\label{fig:distanceplot2}
\end{figure}

\smallskip
\textit{Cut-off versus no cut-off}. Having established the optimal positioning of the repeater, we can now compare the two scenarios: optimised cut-off with middle positioning of the repeater and no cut-off with optimised positioning. We find that in the absence of dark counts the scaling with distance of both schemes is the same, with a small advantage of the cut-off scheme. However, the cut-off is more robust against dark counts. Hence, for imperfect detectors the cut-off allows distributing keys at larger distances. These results can be seen in FIG.~\ref{fig:distanceplotNoDarkCounts} and FIG.~\ref{fig:distanceplot2}, which show the secret-key rate as a function of distance for detectors without and with dark counts, together with the channel capacity of the optical fibre (i.e.~benchmark 1a). We plot the data for the expected and improved parameters ($p_{\textrm{ps}} = p_{\textrm{em}} = 0.6$ and $F_{\textrm{gm}} = 0.97$). 

In FIG.~\ref{fig:distanceplotNoDarkCounts} where we assume no dark counts, we see that for small distances the rate scales approximately with the square root of the transmissivity for both scenarios. That is, they are proportional to the theoretical optimum~\cite{luong2015overcoming} of $\sqrt{\eta_f} = e^{-L/2L_0}$. For sufficiently large distances time-dependent decoherence of the memory $\textrm{QM}_1$ becomes a problem. Both schemes overcome it at the expense of reducing the yield. As a result, the scaling becomes proportional to $\eta_f = e^{-L/L_0}$ for both schemes. In FIG.~\ref{fig:distanceplot2} however we see that the presence of dark counts affects the two schemes quite differently. While for both schemes the effect of dark counts becomes the dominant source of noise after a certain distance, this distance is shorter for the no cut-off scheme than for the scheme with the cut-off. In other words, we see that the cut-off is more robust towards dark counts than the repositioning method. This fact can be explained by noting that shifting the repeater towards Bob increases the losses on Alice's side and as a result makes the Alice-repeater link vulnerable to dark counts. With the cut-off however, the repeater remains in the middle making both of the individual links Alice-repeater and repeater-Bob shorter than the Alice-repeater link in the no cut-off scheme. As a result the setup with the cut-off and with the improved parameters allows us to overcome the channel capacity (1a) more confidently and over larger range of distances, than without the cut-off. 

\begin{figure}[h]
\centering
\includegraphics[scale=0.8,clip,trim = 0mm 25mm 0mm 20mm]{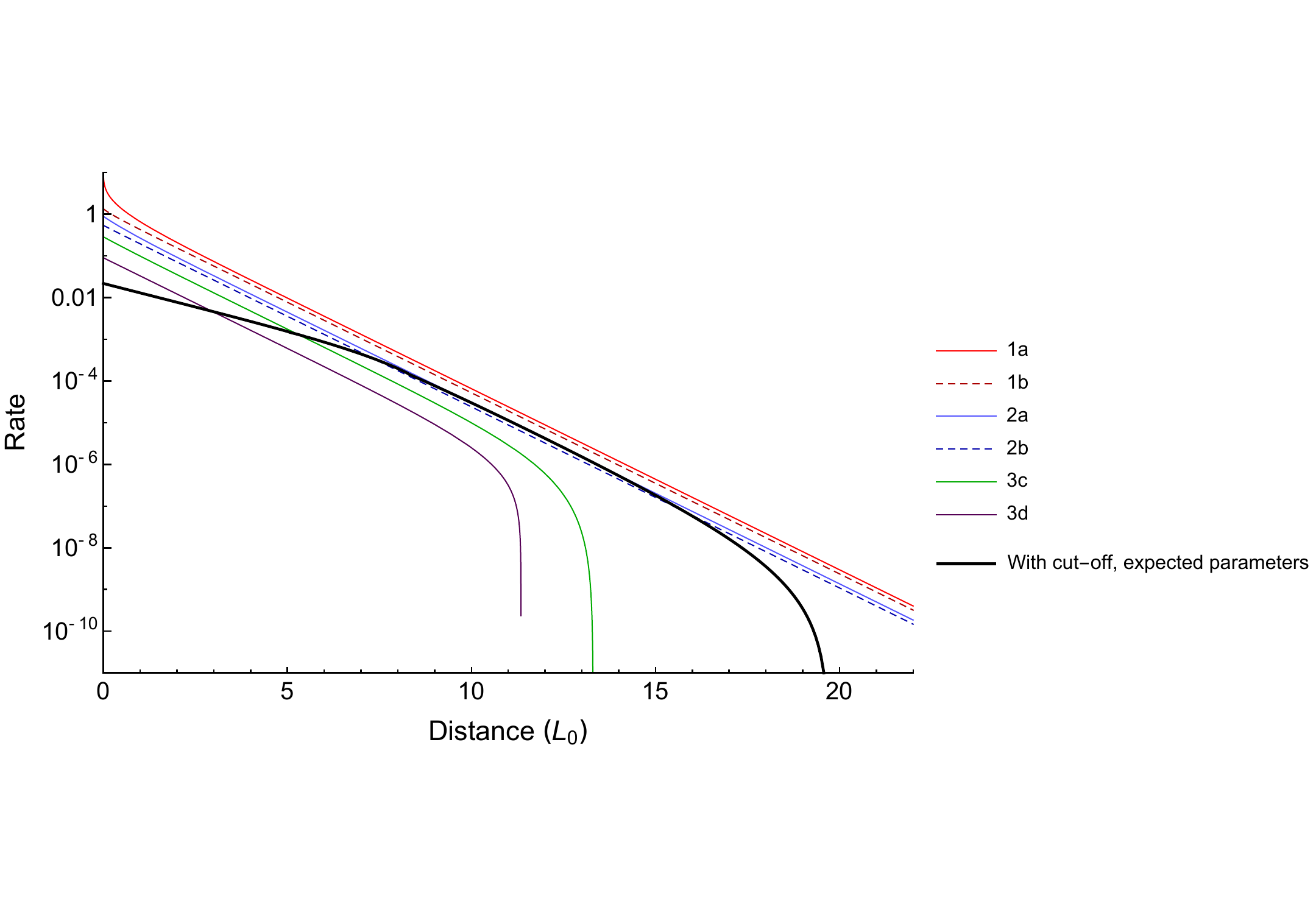}
\caption{Secret-key rate with the quantum repeater implementation for the expected parameters with optimised cut-off as a function of the distance in units of $L_0= 0.542 \text{ km}$. 
The rate is compared to all the benchmarks defined in Table \ref{table:table1}.}
\label{fig:distanceplot3}
\end{figure}

\smallskip
\textit{Comparison with the proposed benchmarks.}
Let us now investigate the secret-key rate achievable with the expected parameters and how it compares with the proposed benchmarks. The comparison is depicted in FIG.~\ref{fig:distanceplot3}. The benchmarks corresponding to direct transmission (3d), the thermal-loss channel (3c) and the pure-loss channel with energy constraint and inclusion of post-selection (2b) are outperformed. The achievable secret-key rate is also very close to the pure-loss channel benchmark with post-selection (2a). The other benchmarks are not overcome but are within experimental reach. 

\begin{figure}[h]
\centerfloat
\includegraphics[scale=0.8,clip,trim = 0mm 27mm 2mm 25mm]{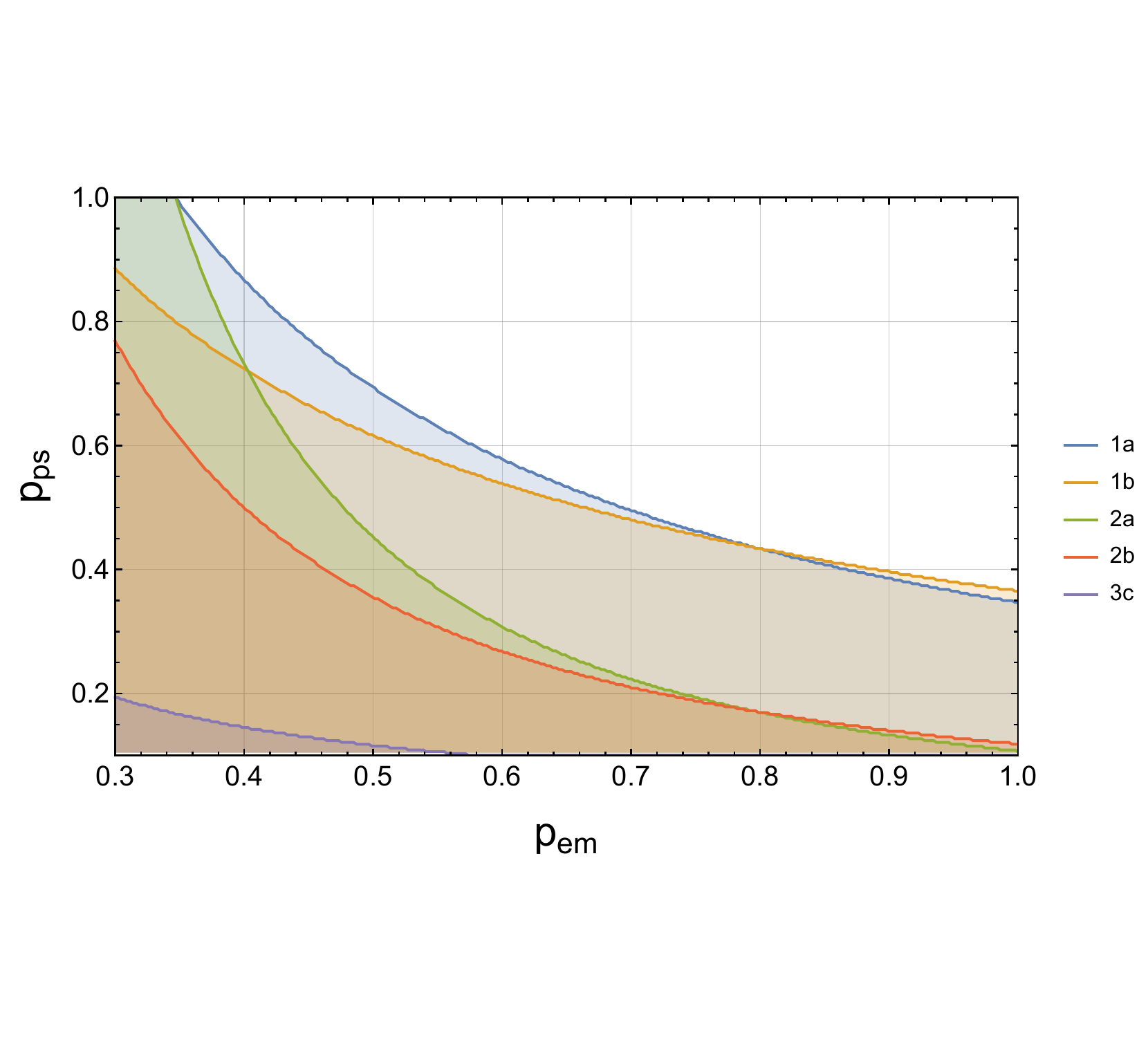}
\caption{Contour plot of regions of $p_{\textrm{em}}$ versus $p_{\textrm{ps}}$ with the expected parameters where the benchmarks listed in Table \ref{table:table1} can be surpassed. The contour lines correspond to the parameters that achieve the corresponding benchmarks while the parameter regimes above the curves allow us to surpass them. The data is plotted for the distance of $9.6L_0$, where $L_0= 0.542 \text{ km}$.}
\label{fig:plot2}
\end{figure}

\begin{figure}[h]
\centerfloat
\includegraphics[scale=0.8,clip,trim = 0mm 27mm 2mm 25mm]{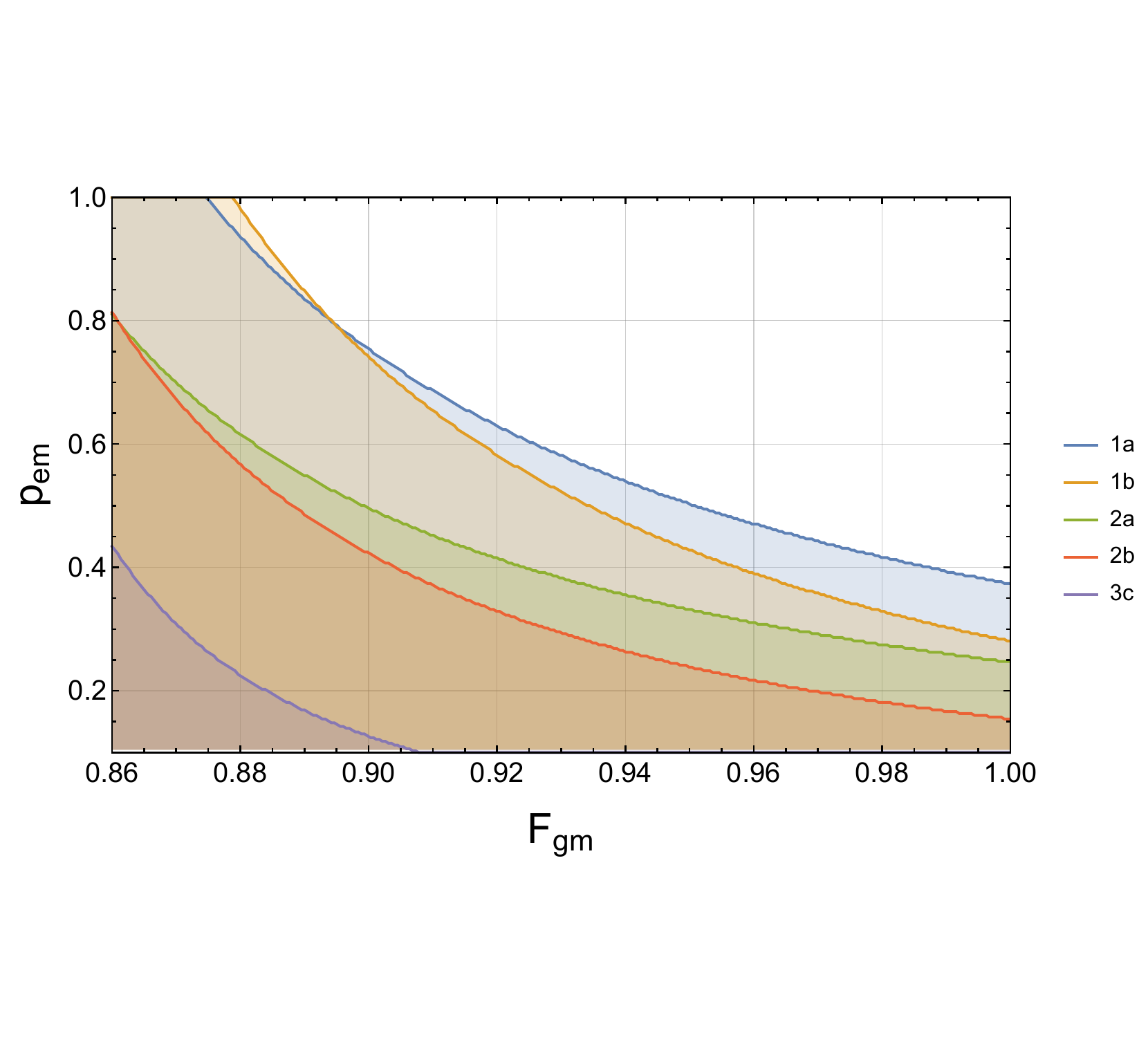}
\caption{Contour plot of regions of $F_{\textrm{gm}}$ versus $p_{\textrm{em}}$ with the expected parameters where the benchmarks listed in Table \ref{table:table1} can be surpassed. The contour lines correspond to the parameters that achieve the corresponding benchmarks while the parameter regimes above the curves allow us to surpass them. The data is plotted for the distance of $9.6L_0$, where $L_0= 0.542 \text{ km}$.}
\label{fig:plot3}
\end{figure}

\smallskip
\textit{Parameter trade-off.}
Let us now give a general overview of how good the improved parameters need to be in order to overcome individual benchmarks. This information is presented on two contour plots. In FIG. \ref{fig:plot2}, we study the parameter regions for which it is possible to beat the benchmarks in Table \ref{table:table1} as a function of $p_{\textrm{ps}}$ and $p_{\textrm{em}}$. A similar plot as a function of $F_{\textrm{gm}}$ and $p_{\textrm{em}}$ can be seen in FIG.~\ref{fig:plot3}. We omit here the direct transmission benchmark which, as we have already seen, can be easily surpassed with the expected parameters. Moreover, we note that the capacity of the thermal channel in the benchmark (3c) goes to zero for very low $p_{\textrm{ps}}$ and $p_{\textrm{em}}$ for which it is still possible to generate key with the quantum repeater. Hence it is trivially easy to beat this benchmark for low $p_{\textrm{ps}}$ and $p_{\textrm{em}}$. In that sense this benchmark is not so interesting in that regime. It is for this reason that this regime is not depicted on the contour plots. In both FIG.~\ref{fig:plot2} and FIG.~\ref{fig:plot3} we observe a crossing between the finite energy benchmarks (1b) and (2b) and their infinite energy counterparts (1a) and (2a) at $p_{\textrm{em}}\approx 0.796$, as discussed in Section~\ref{sec:assessing}.

\begin{figure}[h]
\centering
\includegraphics[scale=0.8,clip,trim = 0mm 25mm 0mm 20mm]{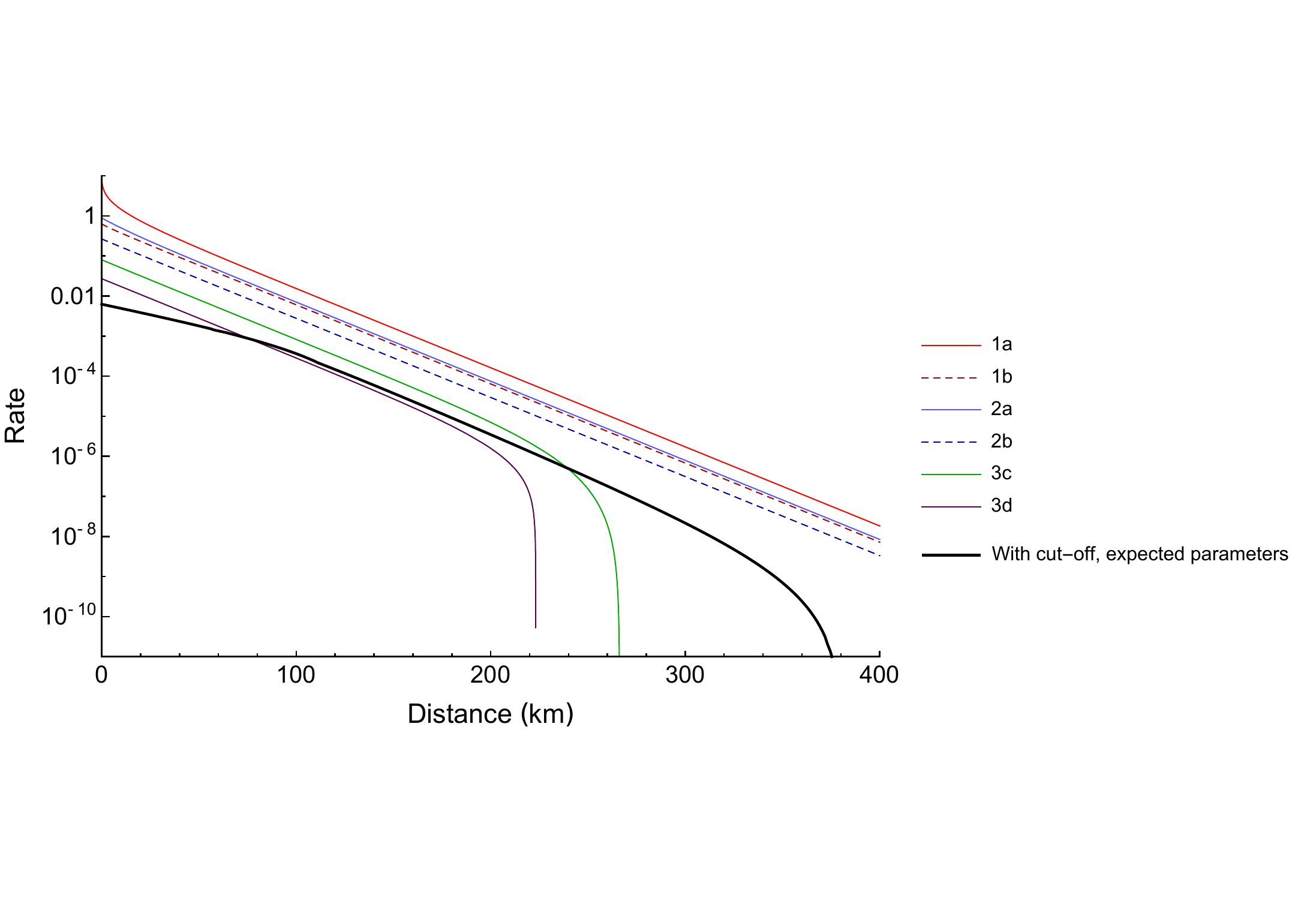}
\caption{Secret-key rate for the telecom channel with $L_0 = 22$ km with the quantum repeater implementation for the expected parameters with optimised cut-off as a function of the distance in units of km. 
The rate is compared to all the benchmarks defined in Table \ref{table:table1}.}
\label{fig:distanceplottelecom}
\end{figure}

\textit{Comparison with the proposed benchmarks for a commonly used telecom channel.}
Let us now again investigate the secret-key rate achievable with the expected parameters and how it compares with the proposed benchmarks, but this time assuming that we have an available channel at the commonly used telecom wavelength with attenuation length $L_0 = 22$ km. Hence in this case the frequency conversion of the emitted light into telecom would be applied. We consider such a conversion process with efficiency of $30 \%$~\cite{zaske2012visible}. This parameter can be added to $p_{\textrm{em}}$ so that we define $p'_{\textrm{em}} = 0.3 \, p_{\textrm{em}}$. We note here that the assumed value of this parameter is a choice based on the specific experimental implementation. However, higher conversion efficiencies are in principle achievable. The comparison is depicted in FIG.~\ref{fig:distanceplottelecom}. We see that for this choice of the direct channel, the benchmarks are more difficult to overcome. In particular only the benchmarks corresponding to direct transmission (3d) and the thermal-loss channel (3c) can be outperformed. The other benchmarks seem to be far from near-term experimental reach.

\section{Conclusions}
\label{sec:conclusions}
\vspace{-2mm}
In this work, we have analysed numerically a realistic quantum repeater implementation for quantum key distribution. We have introduced two methods for improving the rates of the repeater with respect to previous proposals: advantage distillation and the cut-off.
Advantage distillation is a classical post-processing method that increases the secret-key rate at all levels of noise.
The cut-off on the other hand allows for a trade-off between the channel uses required and the secret-key fraction. Utilising the cut-off results in three benefits with respect to the previous scheme for the single sequential quantum repeater~\cite{luong2015overcoming}. Firstly, the cut-off method achieves a higher rate for all distances. Secondly, the protocol is more robust against dark counts, in the sense that non-zero secret key can be generated over larger distances. Finally, the cut-off can be adjusted on the fly, unlike the repositioning of the repeater~\cite{luong2015overcoming}. This is especially convenient in the scenario where the experimental setup might be modified. With the previous scheme for example, improving the coherence times of the memories would lead to a new optimal position. The repositioning of the repeater node would be both costly and time-inefficient, while modifying the cut-off corresponds to a simple change in the programming of the devices.

We note here that one could also use the secret-key rate \emph{per unit time} to assess the performance of a quantum repeater. The secret-key rate per unit time can be calculated by multiplying the secret-key fraction with the inverse of the (average) time it takes to generate a single raw bit between Alice and Bob. This time will depend on the travel time of the photons from the quantum repeater to Alice and Bob, the generation time of the entangled photon-memory pairs and the time it takes to perform the required operations such as the Bell state measurement. To compare the secret-key rate per unit time to the benchmarks, the benchmarks too must then be re-expressed in the secret-key rate per unit time. This can be achieved by multiplying the benchmarks with a fixed emission rate of a photon source~\cite{piparo2017memory}. Note that there is now an ambiguity in the benchmarks, as they depend on the fixed emission rate. Since the emission rate is limited by engineering constraints, the benchmarks are dependent on current technologies and cannot be claimed to be fundamental.

By optimising over the cut-off, we have found realistic parameter regions where it is possible to surpass several different benchmarks including the secret-key capacity. These benchmarks are relevant milestones towards claiming a quantum repeater, and thus form an important step in the creation of the first large-scale quantum networks. 
To make our arguments concrete, we have chosen a specific parameter set induced by some recent experimental results. However, other platforms or technological advances might allow to improve upon our results and predict particularly simple setups for performing the first quantum repeater experiment. For example, our work could be extended by including other types of encoding, such as polarisation encoding, in which case additional depolarising noise in the fibre could become relevant. We leave the investigation of other parameter regimes open. In this respect our model has a very broad functionality, as it allows us to perform efficient optimisation of the secret-key rate over the cut-off for any set of parameters. We achieve this functionality by finding tight analytical bounds for the number of channel uses needed to generate one bit of raw key as a function of the cut-off. Our numerical package is freely available for further exploration~\cite{Note1}.

\section{Acknowledgements}
\vspace{-2mm}
 The authors would like to thank Johannes Borregaard, Suzanne van Dam, Victor Hartong, Peter Humphreys, Thinh Le Phuc, Norbert L{\"u}tkenhaus, Mohsen Razavi and Mark Steudtner for helpful discussions and feedback, and Dmytro Vasylyev for the illustrations of Alice and Bob. This work was supported by the Dutch Organization for Fundamental Research on Matter (FOM), Dutch Technology Foundation (STW), the Netherlands Organization for Scientific Research (NWO) through a VICI grant (RH), a VIDI grant (SW) and the European Research Council through a Starting Grant (RH and SW).

\clearpage

\vspace{5mm}
	      \bibliographystyle{arxiv2}
       \bibliography{library}{}
\onecolumngrid
\appendix
\section{Dark counts}
\label{sec:darkcounts}
In this section we detail the effect of dark counts in the detectors of Alice and Bob on our protocol. In particular, we briefly go over the concept of so-called \emph{squashing models}~\cite{gittsovich2014squashing,beaudry2008squashing}, after which we will be able to calculate the induced depolarising noise. We conclude with explaining how dark counts increase the yield.

Quantum states of light are naturally described by operators on an infinite-dimensional Hilbert space. However, a significant number of optical experiments have been performed where the infinite-dimensional states and operations are approximated by a lower dimensional description. An example of this is where the state of light is assumed to lie within a two-dimensional subspace spanned by the vacuum state and a single-photon excitation. Such an approximation is valid in the sense that the theoretical predictions of measurement statistics correspond accurately to those that are observed experimentally.

However, in cryptographic contexts one usually has to make unconditional statements about the information held by a third party. This third party might be malicious and all-powerful, and her measurement statistics are, by definition, unknown. This implies that there is not necessarily a bound on the information held by a malicious third party, despite the fact that the truncation of the Hilbert space is a good approximation for experimental statistics.\\

Since the theoretical analysis in an infinite-dimensional Hilbert space is difficult, one would prefer to be able to bound the information held by a third party, while at the same time applying a truncation to the finite-dimensional Hilbert space. This can be done if a so-called squashing model exists, which is a way of relating measurements performed on a high-dimensional state to a truncated space. As an approximation we consider here the squashing models for measurements of qubits encoded in the polarisation of photons. In this case squashing models exist for both the fully asymmetric BB84 protocol and the symmetric six-state protocol (with only passive measurements), implying that one can, without loss of generality, perform the fully asymmetric BB84 and symmetric (passive) six-state protocol with photons~\cite{gittsovich2014squashing,beaudry2008squashing}. The squashing model also dictates how multiple clicks in different detectors give rise to noise in the truncated space. In the next section, we discuss how to map the dark counts in the detectors to depolarising noise according to the corresponding squashing model.\\

The parameters typically used to quantify detectors are the dark counts per second and the detection window $t_{\textrm{int}}$, which is the duration of the integration period of the detectors. The number of thermal photons $\overline{n}$ relevant for the thermal benchmark is given by $t_{\textrm{int}}$ times the dark counts per second. Assuming a Poisson distribution of the dark counts, it follows that the probability $p_d$ of getting at least a single dark count click within the time window of awaiting the signal photon is given by $p_d = 1-\exp(-\overline{n})\approx \overline{n}$ for small $\overline{n}$.

The noise caused by the dark counts at Alice's or Bob's detector can then be modelled by a depolarising channel, where the depolarising parameter $\alpha_{A/B}$ depends on the implemented protocol,
\begin{align}
\alpha_{A/B,\textrm{ BB84}} = \frac{p_{\textrm{app}} p_{\textrm{ps}}\eta_{A/B}(1-p_d)}{1-(1-p_{\textrm{app}} p_{\textrm{ps}}\eta_{A/B})(1-p_d)^2}\ , \\
\alpha_{A/B, \textrm{ six-state}} = \frac{p_{\textrm{app}} p_{\textrm{ps}}\eta_{A/B}(1-p_d)^5}{1-(1-p_{\textrm{app}} p_{\textrm{ps}}\eta_{A/B})(1-p_d)^6}\ .
\end{align}
That is, conditioned on a click in at least one of the detectors, Alice or Bob receive the desired state if they receive the signal photon and no other detector was triggered. Due to the squashing map all other events can be mapped onto a maximally mixed state~\cite{gittsovich2014squashing,beaudry2008squashing}. To explain the exponents, we note that the active BB84 protocol requires an optical measurement setup with two detectors, while for the six-state protocol such a measurement setup will consist of six detectors. 

Furthermore, independent of the existence of a squashing map, the dark counts increase the total probability that Alice or Bob gets a click. This probability depends on whether the BB84 or six-state protocol is implemented, and is given by
\begin{align}
p_{A/B,~BB84} = 1 - (1-p_{\textrm{app}} p_{\textrm{ps}}\eta_{A/B})(1-p_d)^2\ , \label{eq:ps1} \\
p_{A/B,~\textrm{six-state}} = 1 - (1-p_{\textrm{app}} p_{\textrm{ps}}\eta_{A/B})(1-p_d)^6\ .
\label{eq:ps2}
\end{align}

\section{Quantum bit error rate}
\label{sec:QBER}
In this Appendix we derive the expressions for the average quantum bit error rate in the $X, \, Y$ and $Z$ basis as a function of the experimental parameters. It is given by
\begin{align}
\langle e_X\rangle	&= \langle e_Y\rangle =\frac{1}{2}-\frac{1}{2}F_{\mathrm{gm}}\alpha_A \alpha_B\left(2F_{\textrm{prep}}-1\right)^2\left\langle e^{-\left(a+b\right)n} \right\rangle\ ,\\
\langle e_Z\rangle	&= \frac{1}{2}-\frac{1}{2}F_{\mathrm{gm}}\alpha_A \alpha_B\langle e^{-b\cdot n}\rangle\ ,
\end{align}
where the average is performed over the geometric distribution with only the first $\nstar$ trials. That is, the average of the exponential $e^{-cn}$ is given by
\begin{align}
\langle e^{-cn} \rangle &= \frac{\sum_{n=1}^{\nstar}p_B\left(1-p_B\right)^{n-1}e^{-cn}}{\sum_{n=1}^{\nstar}p_B\left(1-p_B\right)^{n-1}}\\
&=\frac{p_Be^{-c}}{1-\left(1-p_B\right)^{\nstar}}\frac{1-\left(1-p_B\right)^{\nstar}e^{-c\nstar}}{1-\left(1-p_B\right)e^{-c}}\label{eq:averageexp}\ . \nonumber
\end{align}

To derive these quantum bit error rates, let us firstly define the two-qubit Bell states as
\begin{eqnarray}
\ket{\psi(\san{x},\san{z})} =
\frac{1}{\sqrt{2}}(
\ket{0}\ket{0 + \san{x}} + (-1)^{\san{z}}
\ket{1} \ket{1 + \san{x}\,\,(\textrm{mod} \, 2)}
),
\end{eqnarray}
for $\san{x},\san{z} \in \{0,1\}$.
The noise in the preparation can be modelled as dephasing noise~\cite{togan2010quantum}. The initially generated entangled state between the quantum memory and the state of the photon flying to Alice is then
\begin{align}
\rho_{AR} = F_{\textrm{prep}} \proj{\psi(1,0)} + (1-F_{\textrm{prep}})\proj{\psi(1,1)}\ ,
\end{align}
where $F_{\textrm{prep}}$ is the preparation fidelity of this state. The state in the first quantum memory is now kept stored there. During this time, a second entangled photon-memory is attempted to be generated at the second quantum memory. During these attempts, the state stored in the first quantum memory decoheres through time-dependent dephasing and depolarising noise acting on it. This means that at the time when the second copy is generated, the first copy will have decohered. This second copy will be of the same form as the first one. The decohered first copy is of the form
\begin{align}
\rho'_{AR}	&= F_{T_1} [F_{\textrm{prep}} (F_{T_2} \proj{\psi(1,0)} + (1 - F_{T_2})\proj{\psi(1,1)}) \\
		&+ (1-F_{\textrm{prep}})\left(F_{T_2} \proj{\psi(1,1)} + (1 - F_{T_2}) \proj{\psi(1,0)}\right)] + (1 - F_{T_1}) \frac{\mathbb{I}}{4}\ , \nonumber
\end{align}
where $F_{T_1}, F_{T_2}$ are respectively the depolarising and dephasing parameters due to the decoherence processes on the stored state in the first memory. The fidelity decays exponentially with the number of attempts~\cite{reiserer2016robust} and hence these parameters can be written as
\begin{align}
F_{T_1}	&= e^{-b\cdot n}\ ,\\
F_{T_2}	&= \frac{1 + e^{-a\cdot n}}{2}\ .
\end{align}
Here $n$ is the number of attempts that have been performed on the second memory to successfully generate the repeater-Bob entanglement and the decay rates $a$ and $b$ are defined in the main text.
Hence we can rewrite the state of $\rho'_{AR}$ as
\begin{align}
\rho'_{AR} = F_{T_1} (F_{\textrm{deph},AR} \proj{\psi(1,0)} + (1 - F_{\textrm{deph},AR}) \proj{\psi(1,1)}) + (1 - F_{T_1}) \frac{\mathbb{I}}{4}\ .
\end{align}
where 
\begin{equation}
F_{\textrm{deph},AR} = \frac{1 + (2 F_{\textrm{prep}} - 1) e^{-a n}}{2}\ .
\end{equation}

The entanglement swapping is performed at the two memories at the repeater node. Since the situation is symmetric for all the four measurement outcomes, without loss of generality we can consider the resulting state on $AB$ as if the repeater measured $\ket{\psi(1,0)}$. If a different Bell state was measured, a Pauli rotation could be used to bring the state to this form. The state that we obtain is
\begin{align}
\rho''_{AB}		&= F_{T_1}\big(\left[F_{\textrm{deph},AR} F_{\textrm{prep}} + (1 - F_{\textrm{deph},AR})(1 - F_{\textrm{prep}}) \right] \proj{\psi(1,0)} \\
			& + \left[F_{\textrm{deph},AR}(1 - F_{\textrm{prep}}) + (1 - F_{\textrm{deph},AR}) F_{\textrm{prep}}\right] \proj{\psi(1,1)}\big) + \left(1- F_{T_1}\right) \frac{\mathbb{I}}{4}\ . \nonumber
\end{align}
Finally we note that the operations such as Bell state measurements or any other required gates performed on the memories are also noisy. We will model them by the depolarising channel here~\cite{cramer2015repeated}. The depolarising channel commutes with the dephasing channel. For the two copies of the Bell-diagonal state, it also commutes with the entanglement swapping, in the sense that applying it to one of our memory qubits is mathematically equivalent to applying the same channel to one of the photons flying to Alice or Bob. Hence independently of when exactly in the protocol those gates or measurements on the memories are applied, we can add the resulting depolarisation to the final state shared between Alice and Bob, so that we obtain
\begin{align}
\rho''_{AB}		&= F_{\textrm{gm}}\alpha_A \alpha_B F_{T_1}\big(\left[F_{\textrm{deph},AR} F_{\textrm{prep}} + (1 - F_{\textrm{deph},AR})(1 - F_{\textrm{prep}}) \right] \proj{\psi(1,0)} \\
			& + \left[F_{\textrm{deph},AR}(1 - F_{\textrm{prep}}) + (1 - F_{\textrm{deph},AR}) F_{\textrm{prep}}\right] \proj{\psi(1,1)}\big) + \left(1- F_{\textrm{gm}}\alpha_A \alpha_B F_{T_1}\right) \frac{\mathbb{I}}{4}\ . \nonumber
\end{align}
Here by $F_{\textrm{gm}}$ we denote the product of all the depolarising parameters corresponding to all noisy gates and measurements and $\alpha_{A/B}$ corresponds to the noise caused by the dark counts on Alice's/Bob's side.
From the final state it follows that
\begin{align}
\langle e_X\rangle	&= \langle e_Y\rangle =\frac{1}{2}-\frac{1}{2}F_{\mathrm{gm}}\alpha_A \alpha_B\left(2F_{\textrm{prep}}-1\right)^2\left\langle e^{-\left(a+b\right)n} \right\rangle\ ,\\
\langle e_Z\rangle	&= \frac{1}{2}-\frac{1}{2}F_{\mathrm{gm}}\alpha_A \alpha_B\langle e^{-b\cdot n}\rangle\ .
\end{align}
where the average is over the geometric distribution with only the first $\nstar$ trials. This is due to the fact that, by construction, the state is never allowed to decohere more than $\nstar$ trials.

\section{Comparison with memory-assisted measurement-device-independent QKD schemes}
\label{sec:MDI}

The setup of the proof-of-principle repeater analysed in this paper bears close resemblance to the memory-assisted measurement-device-independent QKD (MA-MDI QKD) setups proposed in~\cite{panayi2014memory}, which were analysed in more detail in the particular context of NV centres in~\cite{piparo2017measurement}. However, in contrast to our focus on key per channel use, these schemes were mostly assessed on their performance of generating key per unit time. In this section, we will briefly discuss these schemes and their advantages and disadvantages in comparison to the scheme analysed in this paper. In particular, we will focus both on their relevance in the context of secret-key generation per channel use, and on the complexity of their experimental implementation.

\begin{center}
\begin{figure}
\centering
\includegraphics[scale=1.05,clip,trim = 25mm 210mm 20mm 10mm]{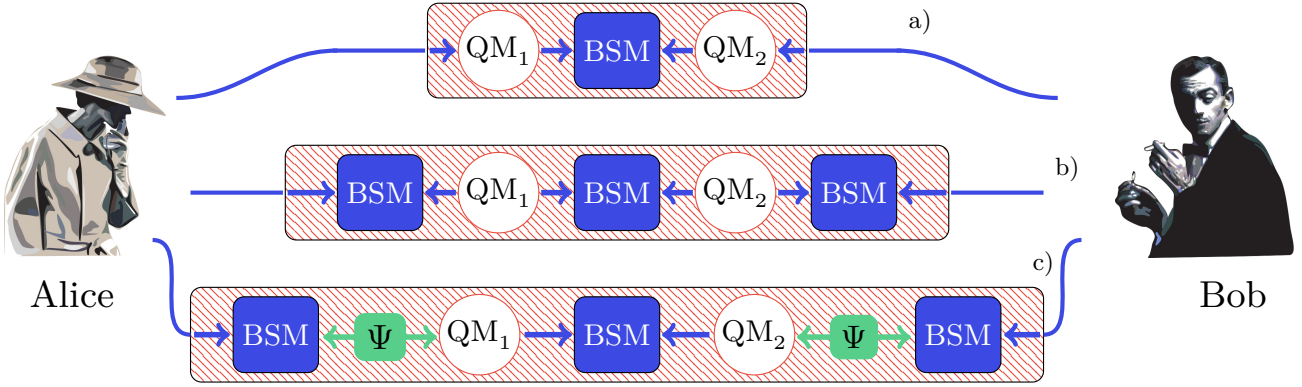}
\caption{Three different setups for the memory-assisted measurement-device-independent quantum key distribution (MA-MDI QKD). Here, BSM stands for Bell state measurement. The first setup a) corresponds to the scheme of MA-MDI QKD with direct heralding. Specifically, the implementation of this setup requires that a photonic state can be transferred into a quantum memory $\textrm{QM}_1$ and $\textrm{QM}_2$ in a heralded fashion. That is, following the transfer attempt, one obtains the information whether the state of the photon emitted at Alice or Bob has been successfully transferred to the desired quantum memory. The second setup b) with indirect heralding is a modification of the first one. Here the requirement of the heralded state transfer has been dropped, at the cost of probabilistic Bell state measurements between two photonic qubits at the outer BSM stations. Finally, the setup in c) is a modification of b), which uses sources of entangled photons ($\Psi$). In this way, the attempt to transfer the quantum state of the photon into the memory is performed only after a successful Bell state measurement.  This can increase the rate per unit time, since writing unto and resetting the memory is a time-consuming process. }
\label{fig:mdi}
\end{figure}
\end{center}

The three schemes that we compare with can be found in Figure~\ref{fig:mdi}. These schemes have the advantage of high expected rate per unit time, since heralding of the successful events now takes place at the repeater. Thus, after a failed attempt the repeater can immediately prepare for receiving another photon, without the need for waiting on any classical communication from Alice and Bob. Furthermore, these schemes are secure against detector side-channel attacks~\cite{lo2012measurement}, since in each scheme there is no quantum information sent from the repeater to Alice or Bob.

However, these advantages, while relevant in practical QKD setups, might not necessarily translate directly in higher secret-key rate per channel use for proof-of-principle repeaters. Moreover, there are experimental challenges that make these MA-MDI QKD schemes more difficult to implement than the sequential quantum repeater that we consider. This is particularly important, since the goal of this paper is to analyse a protocol that would be simple from the implementation perspective, and would have the capability to exceed the benchmarks in Section~\ref{sec:assessing}.

Let us now go over each of these schemes. Firstly, let us consider the first scheme a). This scheme seems to require a similar number of components as our proposed scheme, with the exception that the two detector setups have now been replaced with the sources of BB84 states. The main difficulty with implementing such a scheme lies in the requirement of heralded quantum state transfer from a single photon into the quantum memory. This is a great challenge from the experimental perspective and is not expected to be realised with high fidelity on a significant number of physical platforms in the near future. In systems that utilise cavities this task can be performed, provided that one can realise a low-loss overcoupled cavity with high cooperativity. While such a scenario has been demonstrated experimentally in trapped atoms by achieving the strong coupling regime~\cite{kalb2015heralded}, demonstrating high cooperativity is very challenging in general.

Due to the reasons explained above, scheme b) seems more realistic than scheme a) with the current state-of-the-art technology. However, a larger number of components is needed and the two additional optical Bell state measurements will reduce the rate by a factor of four. In particular, photonic states need to be emitted both from the quantum memories and the BB84 sources. These need to be synchronised such that the Bell state measurements can be performed on both of them. While there is nothing fundamentally challenging with this scheme, it requires larger number of components and is more complicated than the scheme analysed in this paper. Similar conclusions apply to the more complex scheme proposed in c), which adds sources of entangled photons (denoted here by $\Psi$) into the scheme of b). A comparison of the achieved secret-key rate with the secret-key capacity, for a variant of scheme c), has been performed in~\cite{2017arXiv170707814L}.

\section{Secret-key fraction and advantage distillation}
\label{sec:secretkeyfracderiv}
In this section the secret-key fraction formula for the six-state protocol with advantage distillation of~\cite{watanabe2007key} is briefly reviewed. We note here that while the analysis in Appendix~\ref{sec:QBER} has the state $\ket{\psi(1,0)}$ as the target state, here we follow the analysis of~\cite{watanabe2007key} for which $\ket{\psi(0,0)}$ is the target state. This doesn't affect the overall analysis as the final state from Appendix~\ref{sec:QBER} can be rotated locally such that $\ket{\psi(0,0)}$ could be made the target state.
The secret key fraction can be expressed in terms of the Bell coefficients of the Bell diagonal state
\begin{eqnarray}
\rho_{AB} = \sum_{\mathclap{\san{x},\san{z} \in \{0,1\} }}\; P_{\san{XZ}}(\san{x},\san{z})
\ket{\psi(\san{x},\san{z})}\bra{\psi(\san{x},\san{z})}\ .
\end{eqnarray}
Here $P_{\san{XZ}}$ is a probability distribution and we will abbreviate $P_{\san{XZ}}(\san{x}, \san{z})$ as $p_{\san{xz}}$. For the description of the advantage distillation protocol we refer the reader to~\cite{watanabe2007key}. It is shown there that the secret-key fraction can be written as
\begin{eqnarray} 
r_{\text{six-state}} = \frac{1}{3}\max \left[1 - H(P_{\san{XZ}}) + \frac{P_{\bar{\san{X}}}(1)}{2}
h\left( \frac{p_{00} p_{10} + p_{01} p_{11}}{
(p_{00} + p_{01})(p_{10} + p_{11})}\right), 
\frac{P_{\bar{\san{X}}}(0)}{2} (
1 - H(P_{\san{XZ}}^\prime) ) \right] ,
\label{eq-key-rate-vs-error-rate}
\end{eqnarray} 
where
\begin{eqnarray}
P_{\bar{\san{X}}}(0) &=& (p_{00} + p_{01})^2
+ (p_{10} + p_{11})^2\ , \\
P_{\bar{\san{X}}}(1) &=& 2 (p_{00} + p_{01})
(p_{10} + p_{11})\ , \\
P_{\san{XZ}}^\prime(0,0) &=&
\frac{p_{00}^2 + p_{01}^2}{(p_{00} + p_{01})^2 + (p_{10}+p_{11})^2}, \\
P_{\san{XZ}}^\prime(1,0) &=& 
\frac{2 p_{00} p_{01}}{(p_{00} + p_{01})^2 + (p_{10}+p_{11})^2}, \\
P_{\san{XZ}}^\prime(0,1) &=&
\frac{p_{10}^2 + p_{11}^2}{(p_{00} + p_{01})^2 + (p_{10}+p_{11})^2}, \\
P_{\san{XZ}}^\prime(1,1) &=&
\frac{2 p_{10}p_{11} }{(p_{00} + p_{01})^2 + (p_{10}+p_{11})^2} \ ,
\end{eqnarray}
and $H(P_{\san{XZ}})$ is the Shannon entropy of the distribution $P_{\san{XZ}}$. The factor of a third arises from the fact that for a symmetric six-state protocol only a third of the measurements will be performed in the same basis by Alice and Bob.

In our model we only consider depolarising noise and dephasing noise in standard basis. Hence for the six-state protocol the error rates in $X$ and $Y$ basis will be the same. Therefore
\begin{align}
p_{10}+p_{11} &= e_Z\ , \\
p_{01}+p_{11} &= e_{XY}\ , \\
p_{01}+p_{10} &= e_{XY}\ , \\
p_{00} + p_{01} + p_{10} + p_{11} &= 1\ .
\end{align}
Hence
\begin{align}
p_{00} &= 1 - \frac{e_Z}{2} - e_{XY}\ ,\\
p_{01} &=  e_{XY} - \frac{e_Z}{2}\ ,\\
p_{10} &= p_{11} = \frac{e_Z}{2}\ .
\end{align}
And so
\begin{align}
P_{\bar{\san{X}}}(0) &= 1-2e_Z + 2e_Z^2\ ,\\
P_{\bar{\san{X}}}(1) &= 2(1-e_Z)e_Z\ .
\end{align}

\section{Yield}
\label{sec:yieldderiv}
In this Appendix we derive the analytical approximation for the yield
with the cut-off $\nstar$. The yield $Y$ is given by
\begin{equation}
Y = \frac{p_{\textrm{bsm}}}{\mathbb{E}\left[N\right]} = \frac{p_{\textrm{bsm}}}{\mathbb{E}\left[\max(N_A, N_B)\right]}\label{eq:yielddefinition}\ .
\end{equation}
The approximation used for $\mathbb{E}\left[\max(N_A, N_B)\right]$ is
\begin{align}
\mathbb{E}\left[\max(N_A, N_B)\right] \approx  \left\{
\begin{array}{ll}
      \frac{1}{p_A\left(1-\left(1-p_B\right)^{\nstar}\right)} & \frac{1}{p_A}\geq \nstar \\
       \frac{1}{p_A}+\frac{1}{p_B}-\frac{1}{p_A+p_B-p_Ap_B} & \frac{1}{p_A}<\nstar , \\
\end{array} 
\right.
\end{align}
where $p_A$ and $p_B$ are defined in Eq.~\eqref{eq:ps1} for BB84 and in Eq.~\eqref{eq:ps2} for the six-state protocol.
In the rest of this Appendix, we will motivate this approximation by finding tight analytical lower and upper bounds on $\mathbb{E}\left[N\right]$.\\

We note that we consider separately two parameter regimes. One
of them is the regime where on average the dominant number of channel
uses per round is on Alice's side $\left(\frac{1}{p_A}> \nstar \right)$. This corresponds to the high-loss
regime since the number of channel uses per round on Bob's side is
upper bounded by the cut-off. The other regime is the low-loss regime $\left(\frac{1}{p_A}\leq\nstar \right)$.
In this regime we will show that the cut-off does not play any significant role, so that in this regime the formula for the yield with no cut-off~\cite{panayi2014memory, luong2015overcoming} can be used. Moreover, for our derivation to be valid we require an additional constraint to be satisfied, namely $p_B \ge p_A$. This means that we cannot consider scenarios when the repeater is positioned closer to Alice than to Bob. Such a constraint is well-justified since the time-dependent decoherence in quantum memory $\textrm{QM}_1$ would only increase by shifting the repeater towards Alice.

\subsection*{High-loss regime} 
The high-loss regime is the regime where the losses on Alice's side together with the cut-off on Bob's side ensure that the predominant number of channel uses is almost always on Alice's side, i.e.~$\mathbb{E}\left[N\right] = \mathbb{E}\left[\max(N_A, N_B)\right] \approx \mathbb{E}\left[N_A\right]$. This regime is described by the condition $p_A \nstar < 1$. More specifically, as we will show in this section, if

\begin{equation}
\frac{1}{p_A}:= \mu = \beta \nstar, \, \, \beta>1\ ,
\label{eq:approxcond}
\end{equation}
then
\begin{equation}
\mathbb{E}[N_A]  \leq \mathbb{E}\left[N\right] \leq  \left(g_\text{err}(p_A, p_B, \nstar) +1\right) \mathbb{E}[N_A]\label{eq:hlcondition}\ ,
\end{equation}
where $\mathbb{E}[N_A] = \frac{1}{p_A (1-(1-p_B)^{\nstar})}$ (see Eq.~\eqref{eq:NAmean}) and $g_\text{err}(p_A, p_B, \nstar) = \mathcal{O}\left(\frac{1}{\beta^2}\right)$ is a function defined in Eq.~\eqref{eq:gerr1}. This implies that for $\beta$ large enough, $\mathbb{E}\left[N\right]$ can be accurately approximated by $\frac{1}{p_A (1-(1-p_B)^{\nstar})}$.\\

We start the proof of Eq.~\eqref{eq:hlcondition} by first noticing that $\mathbb{E}\left[N_A\right] \leq
\mathbb{E}\left[N\right]$. It is, thus, only necessary to find an
upper bound for $\mathbb{E}\left[N\right]$. Now, let $p(K=k) = (1-p_r)^{k-1} p_r$ be the probability that Bob succeeds in round $k$. Here $p_r = 1-(1 - p_B)^{\nstar}$ is the probability that Bob succeeds in a given round. Then
\begin{equation}
\mathbb{E}\left[N\right]	= \mathbb{E}\left[\max(N_A, N_B)\right] = \sum_{k=1}^\infty p(K=k) \left(\sum_{n_A=k}^\infty \left(\sum_{n_B=(k-1)\nstar +1}^{k\nstar} p(N_A = n_A \land N_B = n_B | K=k) \max(n_A, n_B) \right) \right)\ . \label{eq:AvChann1}
\end{equation}
One can split the sum over $n_A$ in two, depending on whether $n_A$ is greater than $n_B$ or vice versa. We get
\begin{equation} 
\mathbb{E}\left[N\right]= \sum_{k=1}^\infty p(k) \left( \sum_{n_B=(k-1)\nstar +1}^{k\nstar} \left(\sum_{n_A=k}^{n_B} p(n_A \land n_B | k) n_B \right)+\sum_{n_B=(k-1)\nstar +1}^{k\nstar} \left(\sum_{n_A=n_B+1}^\infty p(n_A \land n_B |k) n_A \right) \right)\ ,
\label{eq:AvChann2}
\end{equation}
where $p(k)=p(K=k)$, and $p(n_A \land n_B |k)=p(N_A = n_A \land N_B = n_B |K=k)$. The first term of Eq.~(\ref{eq:AvChann2}) can be upper bounded noticing that $n_B\le k \nstar$, i.e. 
\begin{equation}
\sum_{k=1}^\infty p(k)  \left(\sum_{n_B=(k-1)\nstar +1}^{k\nstar} \left(\sum_{n_A=k}^{n_B} p(n_A \land  n_B | k) n_B \right) \right)\leq \sum_{k=1}^\infty p(k) p\left(N_A \leq N_B \vert K = k\right) k\nstar.
\label{eq:firstterm}
\end{equation}
The second term of Eq.~(\ref{eq:AvChann2}) can be upper bounded in the following way
\begin{align}
\sum_{k=1}^\infty p(k) \left(\sum_{n_B=(k-1)\nstar +1}^{k\nstar} \left(\sum_{n_A=n_B+1}^\infty p(n_A \land n_B |k) n_A \right) \right) &\le \sum_{k=1}^\infty p(k) \left(\sum_{n_A=k}^\infty p(n_A |k) n_A \right)\\
&=\sum_{k=1}^\infty p(k) \sum_{n_A=1}^\infty p(n_A |k) n_A  \\
&=\sum_{n_A=1}^\infty p(n_A) n_A=\mathbb{E}\left[N_A\right]\ .
\label{eq:ENA}
\end{align}
Inputting Eq.~\eqref{eq:firstterm} and Eq.~\eqref{eq:ENA} back into Eq.~(\ref{eq:AvChann2}), we obtain
\begin{equation}
\mathbb{E}[N]\le\left(\frac{\nstar}{ \mathbb{E}[N_A]}\sum_{k=1}^\infty p(k) p\left(N_A \leq N_B \vert k\right) k +1\right) \mathbb{E}[N_A]\ .
\label{eq:AvChann5}
\end{equation}
Let $N_A^i$ be the random variable describing the number of trials on Alice's side in round $i$. Since $p(N_A^i = n_A^i) = (1- p_A)^{n_A^i - 1} p_A$, we clearly have that $\mathbb{E}[N_A^i] = \frac{1}{p_A} = \mu$. Then we note that
\begin{equation}
\mathbb{E}[N_A] = \sum_{k=1}^\infty p(k) \sum_{i=1}^k \sum_{n_A^i = 1}^{\infty} p(n_A^i) n_A^i =  \sum_{k=1}^\infty p(k) \sum_{i=1}^k \mathbb{E}[N_A^i] = \mu \sum_{k=1}^\infty p(k) k = \mathbb{E}[K] \mu = \frac{1}{p_A p_r} = \frac{1}{p_A (1-(1-p_B)^{\nstar})}\ .
\label{eq:NAmean}
\end{equation}
Here, we first express $\mathbb{E}[N_A]$ by calculating the average number of trials in each of the $k$ rounds. Then, we sum the $k$ averages together, and finally, we average over the total number of rounds $k$. Since all the rounds are independent, we replace each $\mathbb{E}[N_A^i]$ by $\mu$ as stated above. By inputting Eq.~\eqref{eq:NAmean} into Eq.~\eqref{eq:AvChann5}, we get
\begin{equation}
\mathbb{E}\left[N_A\right] \leq\mathbb{E}\left[N\right] \leq \left(\frac{1}{ \mathbb{E}[K] \beta}\sum_{k=1}^\infty p(k) p\left(N_A \leq N_B \vert k\right) k +1\right) \mathbb{E}[N_A]\ .
\end{equation}
We now upper bound the $p\left(N_A \leq N_B \vert k\right)$ term. Note that
\begin{equation}
p\left(N_A \leq N_B \vert k\right) = p\left(\sum_{i=1}^{k} N^i_A \le \sum_{i=1}^{k} N^i_B \Big\vert k\right)\ .
\end{equation}
We note that conditioned on $K=k$, we have that $\sum_{i=1}^{k} N^i_B = (k-1)\nstar +N^k_B$. It then follows that
\begin{equation}
p\left(N_A \leq N_B \vert k\right)= p\left( \sum_{i=1}^{k} N^i_A  \le (k-1)\nstar + N^k_B \Big\vert k\right) \leq p\left( \sum_{i=1}^{k} N^i_A  \le k \nstar \Big\vert k\right)\ .
\end{equation}
Condition Eq.~\eqref{eq:approxcond} and $-\sum_{i=1}^{k} N^i_A  \ge - k \nstar$ is equivalent to $k \mu -\sum_{i=1}^{k} N^i_A \ge  k(\beta -1) \nstar$. Hence,
\begin{equation}
p\left(\sum_{i=1}^{k} N^i_A  \le k \nstar \Big\vert k\right) = p\left( k \mu -\sum_{i=1}^{k} N^i_A \ge  k(\beta -1) \nstar \Big\vert k\right)\ .
\end{equation}
We can use the Chernoff bound to upper bound this probability. The Chernoff bound for a random variable $X$ is
\begin{equation}
p(X \ge a) \leq \frac{\mathbb{E} [e^{tX}]}{e^{ta}}, \,\, t >0\ .
\label{eq:chernoff}
\end{equation}
Let $X$ be the sum of $k$ random variables $X_1, X_2, \dots, X_k$, where
\begin{equation}
X_i = \mu - N_{A}^i\ ,
\end{equation}
i.e. $X = \sum_{i=1}^{k} X_i = k \mu - \sum_{i=1}^{k}N_{A}^i.$
From this we can now bound the desired probability. Using
\eqref{eq:chernoff} and $a=k(\beta-1)\nstar$, we obtain the inequality
\begin{align}
p\left( k \mu -\sum_{i=1}^{k}N_{A}^i \ge  k (\beta -1) \nstar \Big\vert k\right)	&\le  \frac{\mathbb{E}\left[\exp\left(t\left(k \mu - \sum_{i=1}^{k}N_A^i\right)\right) \Big\vert k\right]}{e^{t k(\beta-1)\nstar}} \\&= \exp\left[t k\left(\mu - (\beta-1) \nstar\right)\right]\mathbb{E}\left[\Pi_{i=1}^k e^{-t N_A^i}\vert k\right]\ .
\end{align}
Let us now focus on $\mathbb{E}\left[\prod_{i=1}^k e^{-t N_A^i}\vert k\right]$,
\begin{equation}
\mathbb{E}\left[\prod_{i=1}^k e^{-t N_A^i}\vert k\right] = \prod_{i=1}^k \mathbb{E}\left[e^{-t N_A^i}\vert k\right] = \prod_{i=1}^k\left(\sum_{n_A^i = 1}^\infty p_A (1-p_A)^{n_A^i - 1}\, e^{-t n_A^i}\right) = \left(\frac{p_A e^{-t}}{1-(1-p_A)e^{-t}}\right)^k\ .
\end{equation}
Here, after the first equality sign we have used the fact that the random variables $N_A^i$ are independent for different $i$'s. After the second equality we note that all of them have exactly the same geometric distribution over the $k$ rounds. Specifically, it is now important to note that this holds provided that $k$ is the value of $K$ on which we have conditioned, i.e., the success on Bob's side occurs exactly in the $k$'th round. Furthermore, the common ratio $(1-p_A) e^{-t}$ satisfies the convergence condition $\abs{(1-p_A) e^{-t}} <1$ for all $t>0$. This yields
\begin{equation}
p\left(N_A \leq N_B \vert K = k\right) \leq \left(\exp\left[t \left(\frac{1}{p_A} - (\beta-1) \nstar\right)\right]\frac{p_A e^{-t}}{1-(1-p_A)e^{-t}}\right)^{k}\ .
\end{equation}
Let's define the function $f(t)$ as
\begin{equation}
f(t) := \exp\left[t \left(\frac{1}{p_A} - (\beta-1) \nstar\right)\right]\frac{p_A e^{-t}}{1-(1-p_A)e^{-t}}\ .
\end{equation}
This function should be minimised subject to $t>0$ to obtain the tightest bound. A single stationary point is analytically found at
\begin{equation}
t_0 = \ln\left(\frac{(1-p_A) (p_A (\beta-1)\nstar - 1)}{p_A (\beta-1)\nstar + p_A - 1}\right)\ .
\end{equation}
We now want to make sure that $t_0$ always satisfies the condition $t>0$, necessary for applying the Chernoff bound. By condition Eq.~\eqref{eq:approxcond}, the denominator of the above expression inside the logarithm is $p_A (\beta-1)\nstar + p_A - 1 = 1- p_A\nstar + p_A - 1 = p_A (1-\nstar) <0$ as long as $\nstar >1$. From this it follows that $t_0>0$ if and only if
\begin{equation}
(1-p_A) (p_A (\beta-1)\nstar - 1) < p_A (\beta-1)\nstar + p_A - 1\ .
\end{equation}
Clearly this condition is equivalent to $-p_A^2 (\beta-1)\nstar
<0$ which is satisfied for $\beta>1$. This means that $t_0>0$ is
always satisfied. Now note that $f(t=0) = 1$. Moreover, one can also
easily verify that $f'(t=0) = \nstar (1-\beta) < 0$ for $\beta >1$, and that $\lim_{t \rightarrow \infty} f(t) \rightarrow \infty$ as long as $\nstar>1$. These properties of $f(t)$, together with the continuity of $f(t)$, prove that $t=t_0$ corresponds to the global minimum of this function in the regime $t>0$ and that $f(t_0) <1$. Hence, we can now calculate $f(t_0)$ which gives
\begin{equation}
f(t_0) = \left(\frac{(p_A (\beta-1)\nstar-1)(1-p_A)}{p_A(\beta-1)\nstar+p_A-1}\right)^{\frac{1}{p_A} -  (\beta-1)\nstar - 1} (1 - p_A (\beta-1)\nstar)\ .
\end{equation}
This formula can be simplified by substituting the condition Eq.~\eqref{eq:approxcond} to eliminate $\beta$
\begin{equation}
f(t_0) = p_A \nstar \left(\frac{\nstar (1-p_A)}{\nstar-1}\right)^{\nstar-1}\ .
\label{eq:ft0}
\end{equation}
$\mathbb{E}\left[N\right] $ can now be upper bounded by an expression
that depends on $f(t_0)$, that is
\begin{equation}
\mathbb{E}\left[N\right] \leq \left(\frac{1}{ \mathbb{E}[K] \beta}\sum_{k=1}^\infty p(K=k) f(t_0)^k k +1\right) \mathbb{E}[N_A]\ .
\end{equation}
We can now average over the number of rounds $k$,
\begin{equation}
\sum_{k=1}^\infty \frac{p_r}{(1-p_r)} \left[(1-p_r) f(t_0)\right]^k k = \frac{p_r f(t_0)}{\left[1-(1-p_r) f(t_0)\right]^2}\ .
\end{equation}
Moreover, $ \mathbb{E}[K] = \frac{1}{p_r}$ and again removing $\beta$
through condition Eq.~\eqref{eq:approxcond} yields
\begin{equation}
\mathbb{E}\left[N\right] \leq \left(\frac{p_r^2 p_A \nstar f(t_0)}{\left[1-(1-p_r) f(t_0)\right]^2} +1\right) \mathbb{E}[N_A] = \left(\frac{(1 - (1-p_B)^{\nstar})^2 p_A \nstar f(t_0)}{\left[1-(1-p_B)^{\nstar} f(t_0)\right]^2} +1\right) \mathbb{E}[N_A]\ .
\end{equation}
Now by taking the number of channel uses to be $\mathbb{E}\left[N_A\right]$, we can define the relative error $g_\text{err}(p_A, p_B, \nstar)$,
\begin{equation}
g_\text{err}(p_A, p_B, \nstar) := \frac{(1 - (1-p_B)^{\nstar})^2 p_A \nstar f(t_0)}{\left[1-(1-p_B)^{\nstar} f(t_0)\right]^2}\ ,
\label{eq:gerr1}
\end{equation}
with $f(t_0)$ given in Eq.~\eqref{eq:ft0}, so that
\begin{equation}
\mathbb{E}[N_A]  \leq \mathbb{E}\left[N\right] \leq  \left(g_\text{err}(p_A, p_B, \nstar) +1\right) \mathbb{E}[N_A]\ ,
\end{equation}
where the conditions required to satisfy the above formula are $\nstar>1$ and $p_A\nstar<1$. Finally, we can now show how $g_\text{err}(p_A, p_B, \nstar)$ scales with $\beta$. Note that
\begin{equation}
f(t_0) \leq p_A \nstar \left(1 + \frac{1}{\nstar - 1}\right)^{\nstar - 1} \leq  p_A \nstar e\ .
\end{equation}
This together with $f(t_0) <1$ gives
\begin{equation}
g_\text{err}(p_A, p_B, \nstar) < \frac{p_r^2 (p_A \nstar)^2 e}{p_r^2} = \frac{e}{\beta^2}\ .
\end{equation}
Therefore $g_\text{err}(p_A, p_B, \nstar) =
\mathcal{O}\left(\frac{1}{\beta^2}\right)$, implying that the bounds in the high-loss regime are good enough to tightly bound the achieved yield.

\subsection*{Low-loss regime}
Now we consider the complementary low-loss regime characterised by the condition $p_A \nstar \geq 1$. Firstly, since in our protocol there is never any benefit in placing the repeater closer to Alice than to Bob, we also have that $p_B \ge p_A$. This implies that $\frac{1}{p_B} \le \frac{1}{p_A} = \mathbb{E}[N_A^i] \leq \nstar$. This is the regime where the cut-off is large in comparison with the average number of channel uses required to detect a single photon on Bob's side. That is,
\begin{equation}
\frac{\beta'}{p_B} =\nstar, \,\, \nstar \ge \beta'\geq1\ .
\end{equation}
As we will show in this section, in this region we can approximate  $\mathbb{E}\left[N\right] = \mathbb{E}\left[\max(N_A, N_B)\right]$ by $N_{NC}$, where
\begin{equation}
N_{NC} = \frac{1}{p_A} + \frac{1}{p_B} - \frac{1}{p_A + p_B-p_A p_B}\ ,
\end{equation}
is the average number of channel uses in the no cut-off (NC) scenario~\cite{panayi2014memory,luong2015overcoming}. Intuitively, this is because Alice and Bob almost never have to restart due to Bob reaching the cut-off. More specifically, we show that
\begin{equation}
N_{NC}  \leq \mathbb{E}\left[N\right] \leq  \left(\tilde{g}_\text{err}(p_A, p_B, \nstar) +1\right) N_{NC}\label{eq:nccondition}\ ,
\end{equation}
where $\tilde{g}_\text{err}(p_A, p_B, \nstar)$ is defined in Eq.~\eqref{eq:gerr2}. Since $\tilde{g}_\text{err}(p_A, p_B, \nstar) = \mathcal{O}\left(\beta' e^{-\beta'}\right)$, for sufficiently large $\beta'$ the expectation value $\mathbb{E}\left[N\right]$ can be accurately approximated by $N_{NC}$. \\

Here we detail a proof of Eq.~\eqref{eq:nccondition}. We note that the presence of the cut-off increases the number of needed channel uses with respect to the no cut-off scenario, i.e.~$N_{NC} \le \mathbb{E}[N]$. For the upper bound we can write now
\begin{align}
\mathbb{E}\left[N\right]	&= \mathbb{E}\left[\max(N_A, N_B)\right] = \sum_{k=1}^\infty p(K=k) \left(\sum_{n_A=k}^\infty \left(\sum_{n_B=(k-1)\nstar +1}^{k\nstar} p(n_A \land n_B | K=k) \max(n_A, n_B) \right) \right) \\
					&= p(K=1) \sum_{n_B=1}^{\nstar} \sum_{n_A=1}^{\infty} p(n_A | K=1) p(n_B | K=1) \max(n_A, n_B)  \nonumber\\
					&\hspace{3cm} +\sum_{k=2}^\infty p(K=k) \left( \sum_{n_B=(k-1)\nstar +1}^{k\nstar}\left(\sum_{n_A=k}^\infty p(n_A \land n_B | k) \max(n_A, n_B) \right) \right)\ .
\label{eq:AvChannLow2}
\end{align}
In Eq.~\eqref{eq:AvChannLow2} we split the sum over $k$ into two terms, one with $k=1$ and the other with $k>1$. Since the first term has fixed $k=1$, the variables $N_A$ and $N_B$ are independent here (there is only one round in which Bob for sure succeeds, so the value of $n_B$ doesn't affect the value of $n_A$). Moreover, the geometric distribution of $N_B$ is normalised over the interval $[1, \ldots, \nstar]$.
\begin{equation}
\mathbb{E}\left[N\right]\le p(K=1) N_{NC} + \sum_{k=2}^\infty p(K=k) \left(\sum_{n_B=(k-1)\nstar +1}^{k\nstar}\left(\sum_{n_A=k}^\infty p(n_A \land n_B | k) \max(n_A, k \nstar) \right) \right)\ .
\label{eq:AvChannLow3}
\end{equation}
We have upper bounded the first term of Eq.~\eqref{eq:AvChannLow2} by
upper bounding the sum  $\sum_{n_B=1}^{\nstar}$  with $\sum_{n_B=1}^{\infty}$ . In this case the expression after $p(K=1)$ in the first term becomes $N_{NC}$. In the second term we upper bound $n_B$ by $k\nstar$.
Since the second term does not depend on $n_B$ anymore we upper bound
it by removing the constraints on $N_B$ completely from the
probabilities $p(n_A \land n_B | K=k)$, i.e.
\begin{align}
\mathbb{E}\left[N\right]&\le p(K=1) N_{NC} + \sum_{k=2}^\infty p(K=k) \sum_{n_A=k}^\infty p(n_A| K=k) \max(n_A, k \nstar)
\\
&= p(K=1) N_{NC} + \sum_{k=2}^\infty p(K=k) \left(\sum_{n_A=k}^{k\nstar} p(n_A| K=k) k\nstar + \sum_{\mathclap{n_A=k\nstar + 1}}^{\infty}\;p(n_A| K=k) n_A \right)\ ,
\label{eq:AvChannLow5}
\end{align}
where in the last line of Eq.~\eqref{eq:AvChannLow5} we split the second
term into two terms corresponding to the regime where $k\nstar$ is
larger than $n_A$ and vice versa. Since $k\nstar$ does not depend on $n_A$, we upper bound this term by removing the constraints on $n_A$,
\begin{equation}
\mathbb{E}\left[N\right]\le p(K=1) N_{NC} + \sum_{k=2}^\infty p(K=k) k\nstar + \sum_{k=2}^\infty p(K=k) \sum_{n_A=k}^{\infty} p(n_A| K=k) n_A\ .
\label{eq:AvChannLow6}
\end{equation}
Eq.~\eqref{eq:AvChannLow6} can be greatly simplified. We can perform the sum over $n_A$ in the third term obtaining $k\mu$. Then the sums over $k$ can also be easily evaluated so that the right hand side of Eq.~\eqref{eq:AvChannLow6} can be rewritten as
\begin{align}
p(K=1) N_{NC} + \sum_{k=2}^\infty p(K=k) k\nstar + \sum_{k=2}^\infty p(K=k) k \mu &= p(K=1) N_{NC} + (\nstar + \mu)(\mathbb{E}(K) - p(K=1)) \\
					&= \left(p_r + \frac{\nstar + \mu}{N_{NC}} \left(\frac{1}{p_r} - p_r\right)\right)N_{NC} \\
					&= \left(p_r + \left(\frac{\nstar + \mu}{N_{NC}}\right) \left(\frac{1-p_r^2}{p_r} \right)\right)N_{NC}\ .
\label{eq:AvChannLow9}
\end{align}
 Hence we have that
\begin{equation}
N_{NC}  \leq \mathbb{E}\left[N\right] \leq  \left(\tilde{g}_\text{err}(p_A, p_B, \nstar) +1\right) N_{NC}\ ,\label{eq:Nnc}
\end{equation}
where $\tilde{g}_\text{err}(p_A, p_B, \nstar)$ is defined as
\begin{equation}
\tilde{g}_\text{err}(p_A, p_B, \nstar): = (1-p_B)^{\nstar} \left[\left(\frac{\nstar + \mu}{N_{NC}}\right)\left(\frac{2-(1-p_B)^{\nstar}}{1-(1-p_B)^{\nstar}}\right) - 1\right]\ .
\label{eq:gerr2}
\end{equation}
We now show that $\tilde{g}_\text{err}(p_A, p_B, \nstar)$ is small
compared to the other quantities in Eq.~\eqref{eq:Nnc}.
Observe that 
\begin{equation}
(1-p_B)^{\nstar} = \left(1-\frac{\beta'}{\nstar}\right)^{\nstar} \le e^{-\beta'}\ .
\end{equation}
From Eq.~\eqref{eq:gerr2} it follows that
\begin{equation}
\tilde{g}_\text{err}(p_A, p_B, \nstar) \le e^{-\beta'}\left[\frac{\nstar + \frac{1}{p_A}}{N_{NC}}\left(\frac{2}{1-e^{-\beta'}}\right) -1 \right]\ .
\end{equation}
To upper bound the relative error, we start by upper bounding the first term inside the brackets, namely
\begin{equation}
\frac{\nstar + \frac{1}{p_A}}{N_{NC}}= \frac{\nstar + \frac{1}{p_A}}{\frac{1}{p_A} + \frac{1}{p_B} - \frac{1}{p_A + p_B-p_A p_B}} \le \frac{\nstar + \frac{1}{p_A}}{\frac{1}{p_A} + \frac{1}{p_B} - \frac{1}{p_A + p_B-p_A}} = p_A \nstar + 1\ .
\end{equation}
$\tilde{g}_\text{err}(p_A, p_B, \nstar)$, then, is upper bounded by
\begin{align}
\tilde{g}_\text{err}(p_A, p_B, \nstar)	&\le e^{-\beta'}\left[(p_A \nstar + 1)\left(\frac{2}{1-e^{-\beta'}}\right) -1 \right] \\
							&= \frac{e^{-\beta'}}{1 - e^{-\beta'}} (2p_A\nstar + 1 + e^{-\beta'}) \\
							&\le \frac{e^{-\beta'}}{1 - e^{-\beta'}} (2\beta' + 1 + e^{-\beta'}) \\
							&= e^{-\beta'}\left(\frac{2\beta'}{1-e^{-\beta'}} + \coth\left(\frac{\beta'}{2}\right)\right) \\
							&< e^{-\beta'}\left(\frac{2\beta'}{1-e^{-1}} + \coth\left(\frac{1}{2}\right)\right) \\
							& < e^{-\beta'}\coth\left(\frac{1}{2}\right)\left(2\beta' + 1\right) \\
							&< 3\coth\left(\frac{1}{2}\right)\beta' e^{-\beta'}\ .
\end{align}
Therefore $\tilde{g}_\text{err}(p_A, p_B, \nstar) = \mathcal{O}\left(\beta' e^{-\beta'}\right)$.

\end{document}